\documentclass[useAMS,usenatbib,usedcolumn]{mn2e}

\usepackage{graphicx}
\usepackage{amsmath}
\usepackage{longtable}
\usepackage{txfonts}
\usepackage{natbib}
\usepackage{multirow}
\usepackage{layout}
\usepackage{url}
\usepackage{nicefrac}
\usepackage{keyval}
\usepackage{wasysym}
\usepackage{amsfonts}
\usepackage{rotating}
\usepackage[T1]{fontenc}
\usepackage{aecompl}

\title[Star-formation in Hickson Compact Groups]{Star-formation properties of Hickson Compact Groups based on deep H$\alpha$ imaging}
\author[Eigenthaler, Ploeckinger, Verdugo \& Ziegler]
{Paul Eigenthaler$^{1}$, Sylvia Ploeckinger$^{2,3}$, Miguel Verdugo$^{2}$, Bodo Ziegler$^{2}$\\
$^1$  Instituto de Astronom\'{i}a y Astrof\'{i}sica, Pontificia Universidad Cat\'{o}lica de Chile, Av.\ Vicu\~{n}a Mackenna 4860, Santiago, Chile\\
$^2$  Institut f\"ur Astrophysik, Universit\"at Wien, T\"urkenschanzstra\ss e 17, 1180 Wien, Austria\\
$^3$  Leiden Observatory, Leiden University, PO Box 9513, NL-2300 RA Leiden, the Netherlands \\
email: eigenth@astro.puc.cl}

\begin{document}

\date{xxx}

\pagerange{\pageref{firstpage}--\pageref{lastpage}} \pubyear{2015}

\maketitle

\label{firstpage}

\begin{abstract}
We present deep H$\alpha$  imaging of seven Hickson Compact Groups (HCGs)  using the 4.1m Southern Astrophysics Research (SOAR)  Telescope. The high spatial resolution of the  observations allow us to
study both  the integrated star-formation  properties of the main  galaxies as well  as the 2D  distribution of star-forming  knots in the  faint tidal arms that  form during interactions  between the
individual galaxies. We derive  star-formation rates and stellar masses for group  members and discuss their position relative  to the main sequence of star-forming galaxies.  Despite the existence of
tidal features within the galaxy groups, we do not find any indication for enhanced star-formation  in the selected sample of HCGs. We study azimuthally averaged H$\alpha$ profiles of the galaxy disks
and compare them  with the $g'$ and $r'$ surface brightness  profiles. We do not find any  truncated galaxy disks but reveal that  more massive galaxies show a higher  light concentration in H$\alpha$
than less massive ones. We also see that galaxies that show a high light concentration  in $r'$, show a systematic higher light concentration in H$\alpha$. TDG candidates have been previously detected
in $R$-band images for 2  groups in our sample but we find that most  of them are likely background objects as they  do not show any emission in H$\alpha$. We present a  new tidal dwarf galaxy (TDG)
candidate at the tip of the tidal tail in HCG 91.
\end{abstract}


\section{Introduction}
A key  question in  ongoing extragalactic research  is to understand  to what  extent galaxy properties  are determined  by initial conditions  (nature) or driven  by environmental  effects (nurture).
Observations in  the last  few decades  have revealed  that the  evolution of  galaxies is  indeed strongly  dependent on  their environment.  In fact,  most galaxies  are not  isolated, but  found in
gravitationally bound  aggregates, such as  groups or  clusters \citep{tully87,abell89,nolthenius93}. Hence,  to fully understand  galaxy evolution and  assess the  relative importance of  nature vs.\
nurture, it is essential to investigate the various physical processes in these different  environments modifying galaxy properties such as star-formation rate (SFR), morphology, kinematics and color.
In this context, compact  groups of galaxies provide an ideal laboratory  to study the effects of galaxy-galaxy  interactions on the evolution of individual group  members. \citet{hickson82} defined a
catalogue of 100 compact groups (hereafter Hickson Compact Groups, HCGs) with 451 galaxies in total.  Although some galaxies in this sample turned out to be chance projections along the line-of-sight,
not physically bound to the corresponding group, the \citet{hickson82} sample is still the most detailed studied  sample of compact groups in the literature due to the vicinity of member galaxies at a
median distance  of $89\,h^{-1}\,$Mpc \citep{hickson92}. The  comparatively low velocity  dispersions in the order  of $200\,$km s$^{-1}$ and  the high spatial  density with a median  projected galaxy
separation of  only $39\,h^{-1}\,$kpc \citep{hickson92}  make HCGs the perfect  targets to investigate  ongoing galaxy transformations  due to galaxy-galaxy interactions.  Based on their  high spatial
density and low velocity  dispersions, crossing times for HCGs are short ($t_{\rm  cr}\le 0.02\,H_0^{-1}$; \citealt{diaferio94}) making their mere existence puzzling.  In fact, dark matter simulations
have shown that HCGs should merge into a single massive galaxy within a Gyr \citep{barnes85,bode93},  being possibly the precursor of so-called fossil groups. One explanation for the existence of HCGs
is that they  have just recently formed and started  to interact on short timescales. \citet{claudia94}  studied the morphology of galaxies in  92 HCGs with at least three  accordant members and found
that 43\% of all galaxies in their sample show morphological or kinematical distortions indicative of  interations or mergers while \citet{hickson82} noted that the spiral fraction in HCGs is a factor
of two lower than in field galaxies. These observations hint towards a scenario where  galaxies in compact groups undergo frequent interactions and mergers, depleting late-type galaxies while building
up a higher early-type fraction instead.  During this very dynamic stage it is possible to study  the impact of the high density environment on the evolution of  the individual group members, which is
not yet fully understood. While interactions in galaxy pairs typically increase the SFR, e.g.\  due to a compressing tidal field \citep{renaud09}, no statistically significant increase in the specific
star-formation rates (sSFR) of galaxies in HCGs  has been found \citep{bitsakis10}. A possible explanation for the reduced SFR in HCGs is  efficient gas stripping through ongoing tidal interactions, a
process  that is  typical for  the  compact group  environment. It  has been  shown  that other  possible mechanisms  to  suppress or  even quench  star-formation  in  HCGs are  shocks and  turbulence
\citep{alatalo}. \citet{cluver13} studied  a sample of 74 galaxies  in 23 Hickson Compact Groups  (HCGs) and find evidence  for enhanced warm H$_2$ emission  in $\sim 20$\% of these  galaxies, most of
which lie in the optical green valley between  the blue cloud and red sequence. This emission has been associated with the dissipation of mechanical  energy caused by a large-scale shock, induced when
one group member collides at  high velocity with tidal debris in the intragroup medium. Hence,  shock excitation or turbulent heating are likely responsible for the  enhanced H$_2$ emission in compact
group galaxies and the suppression  of star-formation. Other sources of heating like UV or X-ray  excitation from star-formation or AGN activity are insufficient to  account for the observed emission.
\citet{verdesmontenegro01} found a mean HI deficiency of 40  \% for 48 HCGs and noted that groups with a higher early-type fraction or more compact  systems with larger velocity dispersions tend to be
more HI deficient. \citet{desjardins13} studied  9 HCGs in X-rays and have shown that HCGs with  a higher X-ray luminosity show a lower sSFR. These observations  suggest an evolutionary scenario where
the amount of detected HI  in galaxies, and hence the sSFR, decreases with evolution by  continuous tidal stripping and/or gas heating to X-ray wavelengths. These  distortions should also be reflected
in the observed  H$\alpha$ profiles of the galaxy disks,  which are expected to be truncated  based on the frequent interactions. Besides  the distortion of gas disks within  the galaxies, the ongoing
interactions can also efficiently remove gas from its host galaxies, forming long  filamentary structures and bridges \citep{iglesiasparamovilchez01,serra12}. These arms can become the birthplace of a
new generation of actively star-forming star clusters  and so-called tidal dwarf galaxies (TDGs; see \citealt{duc12} and references therein). \citet{hunsberger96}  listed 47 TDG candidates in 15 tidal
features within  HCGs and \citet{hunsberger98}  showed that the faint  end of the  luminosity function is  indeed enhanced in HCGs,  possibly due to  the efficient formation  of TDGs. In this  work we
continue these studies  and investigate the spatial distribution  of ongoing star-formation in a  sample of 7 HCGs  in the southern hemisphere by  performing deep H$\alpha$ imaging with  the 4.1m SOAR
telescope. Specifically we compare the  integrated SFRs of group members with the main  sequence of star-forming galaxies and analyze the azimuthally-averaged  H$\alpha$ surface brightness profiles of
the interacting galaxies. By comparing  these profiles with the corresponding surface brightness profiles of the  stellar component, we can reveal any distortions in the  gas component of the galaxies
caused by environmental effects such as tidal stripping. The deep observations also allow us to  detect actively star-forming TDG candidates within the tidal tails of the interacting galaxies. For two
groups in our sample, HCG26 and HCG96, we compare  our H$\alpha$ maps with the work of \citet{hunsberger96}, who claim to detect new TDG candidates in  these systems based on deep $R$ band images. All
groups in our sample have been selected based on their recession velocities ($cz\sim8000\,$km  s$^{-1}$) and morphological appearance, i.e.\ all groups show strong interactions between their brightest
galaxies. \\

The paper  is organised as follows.  In Sect.\ 2 observations,  data reduction and photometric  measurements are described while  Sect.\ 3 focuses on  data analysis and presents  all obtained results.
Section 4  discusses individual groups  and compares our  findings with studies  from the literature.  Conclusions are given in  Sect.\ 5. Magnitudes  presented in this  work are AB  magnitudes except
otherwise stated. Throughout the paper, the standard $\Lambda$CDM cosmology with $\Omega_{\rm{M}} = 0.3$, $\Omega_{\Lambda} = 0.7$, and a Hubble constant of $H_{0}=70$ km s$^{-1}$ is used.

\section{Observations, Data Reduction and photometry} 
\begin{figure}
\includegraphics[width=\columnwidth]{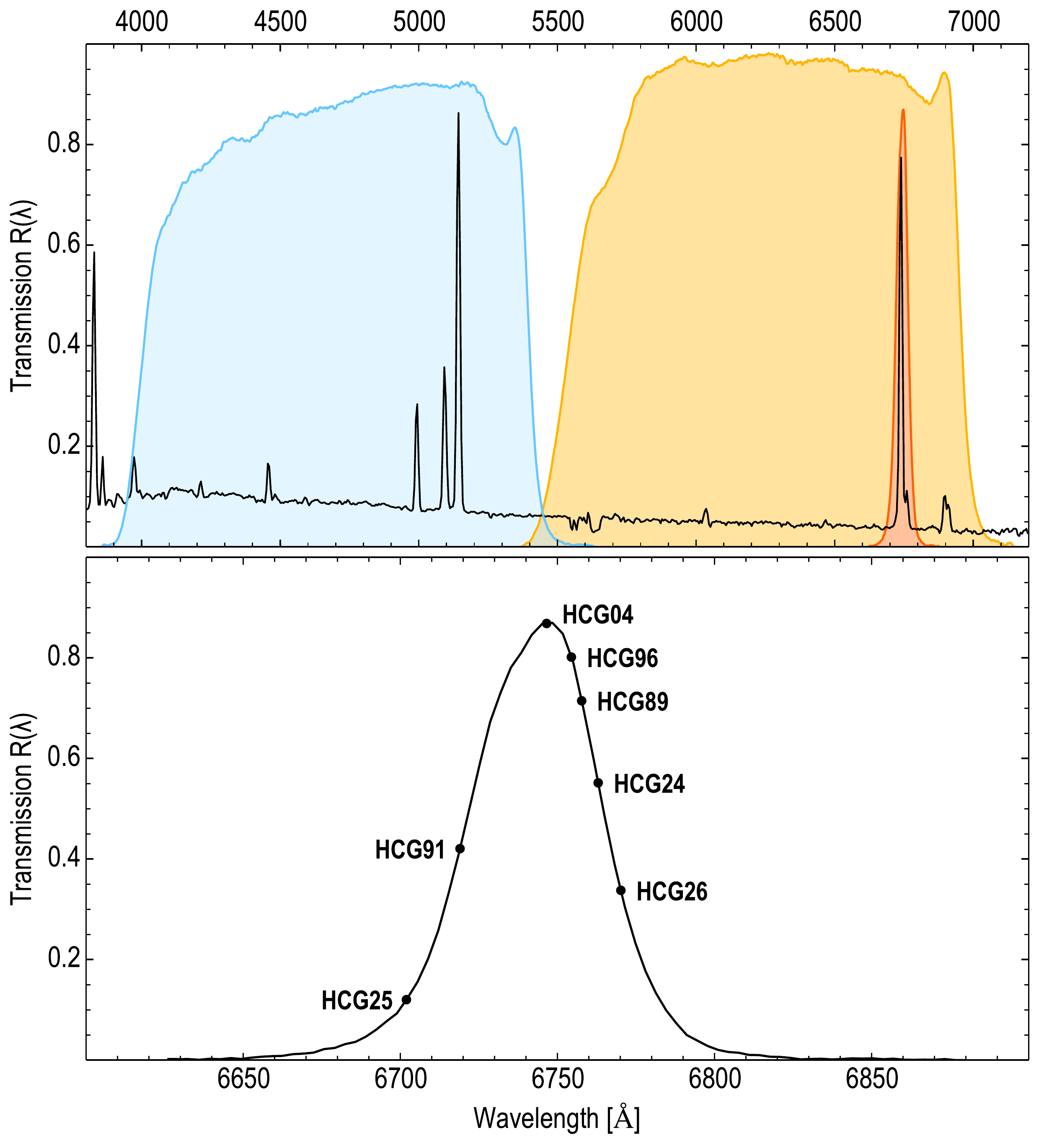}
\caption{\label{transmission}\emph{Upper panel:} Transmission curves of the used filters. Blue: SDSS $g'$, yellow:  SDSS $r'$, orange: S{\scriptsize II} narrowband. The black line indicates a spectral
template of a starburst galaxy redshifted  to the group distance of HCG04, part of our sample. The S{\scriptsize  II} narrowband filter encompasses the [N{\scriptsize II}]$-$H$\alpha$$-$[N{\scriptsize
II}] triplet.  The spectral  template was  taken from  \citet{calzetti94}. \emph{Lower  panel:} Filled  cirlces mark the  positions of  the redshifted  H$\alpha$ line  in our  group sample  within the
S{\scriptsize II} narrowband filter.}
\end{figure}

We carried out deep imaging at the 4.1m  SOAR telescope during one night in 2013$-$October 10 in visitor mode, utilizing the Goodman spectrograph  in imaging mode. The Goodman spectrograph consists of
$4096 \times 4096$  pixels with a pixel size  of $15\mu$m yielding a spatial  scale of 0.15 arcsec pixel$^{-1}$  in $1 \times 1$ imaging  mode. This configuration corresponds to  an effective circular
field-of-view (FOV) of $\sim7.2$ arcmin in diameter due  to the assembly of the spectrograph slit changer. We chose Goodman instead of the SOAR Optical  Imager (SOI) due to the slightly larger FOV and
the higher throughput in  all observed passbands. At the typical redshift  of our group sample ($z\sim0.03$), the observed  FOV corresponds to a circular, physical area  of $\sim250\,$kpc in diameter,
covering the bright, central interacting galaxies of all our HCGs with only one pointing. The seeing  ranged from 0.6 to 1.0 arcsec during the night, typically measuring 0.8 arcsec yielding a physical
resolution  of $\sim500\,$pc  (see Table  \ref{sample}). To  construct H$\alpha$  maps for  our sample  of HCGs,  we obtained  deep images  in the  S{\scriptsize II}  narrowband filter  ($\lambda_{\rm
eff}=6743$\AA, FWHM$=67$\AA), matching  the position of the H$\alpha$  line at the redshift of  our targets. In addition,  we took broadband images in  SDSS $r'$, to estimate the  continuum around the
H$\alpha$ line, and in SDSS  $g'$ to obtain $g'-r'$ color information for all galaxies in  our FOV. Figure \ref{transmission} shows the corresponding filter transmission curves,  as well as a spectral
template of a starburst galaxy redshifted to the group distance of HCG04, part of our sample.

\begin{table*}
\begin{center}
\caption{Sample of Hickson Compact Groups observed at the SOAR Telescope.}
\begin{tabular}{cccccccccc}
\hline                                                                                                                                                                                          
   group   &   $\alpha_{2000}$   &  $\delta_{2000}$    &       $z\,^a$    &  $v$ [km s$^{-1}$]$\,^a$  & $D$ [Mpc]$\,^b$  & $g'\,^{c}$  &   $r'\,^{c}$  &   S{\scriptsize II}$\,^{c}$    &  resolution$\,^{d}$   \\
\hline                                                                                                                                                                                            
   HCG04   &   00 34 16.0        &       $-$21 26 48   &      0.0280      &            8277           &                123            &   360s     &    580s      &     1800s                      &        450pc          \\             
   HCG24   &   03 20 18.9        &       $-$10 51 53   &      0.0305      &            9004           &                134            &   360s     &    610s      &     1800s                      &        488pc          \\             
   HCG25   &   03 20 43.7        &       $-$01 03 07   &      0.0212      &            6288           &      \phantom{1}92            &   360s     &    480s      &     1800s                      &        343pc          \\             
   HCG26   &   03 21 54.2        &       $-$13 38 45   &      0.0316      &            9324           &                139            &   360s     &    480s      &     1800s                      &        505pc          \\             
   HCG89   &   21 20 10.8        &       $-$03 54 32   &      0.0297      &            8772           &                130            &   360s     &    480s      &     1800s                      &        476pc          \\             
   HCG91   &   22 09 12.4        &       $-$27 46 33   &      0.0238      &            7050           &                104            &   160s     &    430s      &     1150s                      &        384pc          \\             
   HCG96   &   23 27 58.3        &       $+$08 46 27   &      0.0292      &            8626           &                128            &   320s     &    450s      &     1800s                      &        468pc          \\  
\hline
\end{tabular}
\label{sample}
\end{center}
{\bf Notes:} $^{a}\,$Redshifts and recession velocities taken from SIMBAD.
             $^{b}\,$Luminosity distances have been computed with the Cosmology Calculator \citep{cosmology}.
		   $^{c}\,$Values show total exposure times for filters $g'$, $r'$ and S{\scriptsize II}.
		   $^{d}\,$Spatial resolution in pc assuming an average seeing of 0.8 arcsec.  
\end{table*}

\subsection{Data Reduction} 
For every  compact group, we  split the observations into  individual dithered exposures  to allow cosmic  ray removal and  correct for bad  pixels. Table 1 summarizes  the observations and  lists all
observed targets. We took bias and domeflats in  the afternoon and skyflats during twilight. To calibrate our data photometrically we observed two standard star  fields in $g'$ and $r'$ in SDSS Stripe
82 \citep{stripe82},  one at  the beginning,  the other  at the end  of the  night. To  calibrate the  S{\scriptsize II} narrowband  filter, we  observed the  spectrophotometric standard  star LTT1020
\citep{hamuy92,hamuy94} with this  filter in the middle  of the night. Data reduction  was carried out within {\sevensize  IRAF}. To remove the  instrument signature from our science  frames, we first
created a masterbias  to remove the bias  level including any two-dimensional  structures. Masterflats were created in  all filters by average  combining the bias-corrected individual  flat frames and
normalizing the combined flats by  the mode of the pixel distribution. Science frames were  then flat-fielded and aligned spatially with the {\sevensize IRAF} task  {\sevensize imalign} to correct for
the 10  arcsec dither, previously introduced  to avoid bad pixels.  For all frames,  the alignment was checked  by eye, comparing  the individual and co-added  images. Before combining our  images, we
corrected the reduced  and aligned science frames  for cosmic rays utilizing  the Laplacian edge detection  technique {\sevensize L.A.COSMIC} \citep{vandokkum}.  This method proved to  be efficient in
detecting and  removing cosmic rays from  our images. Once  all individual images had  been cleaned from  cosmics, we normalized all  frames by the  corresponding exposure times to  obtain countrates.
Subsequently, we average  combined all individual frames  rejecting $3\sigma$ outliers to  account for any remaining  cosmic ray artefacts not  detected via {\sevensize L.A.COSMIC}.  We calibrated our
science  frames astrometrically  within the  {\sevensize  STARLINK GAIA}\footnote{\url{http://starlink.jach.hawaii.edu}}  package  \citep{gaia} by  matching sources  from  the 2MASS  catalog with  the
corresponding sources in our  images until a deviation in the order of  a negligible pixel fraction was achieved. The  final calibration was checked by overplotting the 2MASS  catalog on the resulting
astrometrically calibrated science frames.  We subtracted the sky from our images by  computing the mode of the pixel distribution  in each frame and subtracting it. This method  proved to be adequate
for our data since no spatial trends in the sky background were found in the observed  FOV. Finally, the co-added frames from all three filters were aligned spatially with {\sevensize imalign} so that
the position of the observed sources match in each passband.

\subsection{Photometric calibration\label{photcal}}

Photometric calibration was performed by computing photometric zeropoints via

\begin{equation}
\label{photcalibration}
{m_{{\rm{SDSS}}}} = {m_{{\rm{inst}}}} + {Z_P} - kX =  - 2.5\log \left( {{\rm{ADU/sec}}} \right) + {Z_P} - kX
\end{equation}

where ${m_{{\rm{inst}}}} = - 2.5\log \left(  {{\rm{ADU/sec}}} \right)$ is the measured instrumental magnitude, ${m_{{\rm{SDSS}}}}$ the SDSS magnitude on  the AB system (\citealt{okegunn83}), $Z_P$ the
photometric zeropoint, $k$ the extincion coefficient, and  $X$ the airmass during exposure. To determine photometric zeropoints we measured instrumental magnitudes in  SDSS $g'$ and $r'$ for all stars
from the  \citet{stripe82} catalog falling  in our two Stripe82  fields. We corrected  for airmass considering  extinction coefficents $k_{g'}=0.12$ and  $k_{r'}=0.11$ determined for  Cerro Pach\'{o}n
\citep{ryder06}. Following  equation \ref{photcalibration}  we compute  zeropoints $Z_{P,g'}=26.56  \pm 0.12$  and $Z_{P,r'}=26.53\pm0.08$.  To calibrate  the S{\scriptsize  II} narrowband  filter, we
integrated the  flux of  the spectrophotometric  standard star  $f(\lambda)$ within  the filter transmission  $R_{\rm{SII}}(\lambda)$ and  compared this  value with  the extinction  corrected measured
countrates $c_{{\rm{SII}}}$. To account for extinction in S{\scriptsize II} we utilized the $r'$ band extinction coefficient. Following this procedure, we derive a zeropoint

\begin{equation}
\label{halphacalibration1}
\kappa  = \frac{{\int{f(\lambda)}{R_{{\rm{SII}}}}(\lambda)d\lambda}}{{{c_{{\rm{SII}}}}[{\rm{ADU/sec}}]}}
\end{equation}
 
of $\kappa=8.484\times10^{-17}$ ergs s$^{-1}$cm$^{-2}$ ADU$^{-1}$.

\subsection{Image analysis}
To measure  the net-countrate in the  H$\alpha+$[N{\scriptsize II}] lines,  we first created  continuum maps by scaling  the observed countrates  in the $r'$  band to the S{\scriptsize  II} narrowband
filter. Then the net  H$\alpha+$[N{\scriptsize II}] counts are given by ${c_{{\rm{H}}\alpha+[{\rm{N  {\scriptsize II}]}}}} = {c_{{\rm{SII}}}} - n{c_{r'}}$, where $c_{r'}$  and $c_{{\rm{SII}}}$ are the
countrates in the  $r'$ band and narrowband  frames, respectively. We estimated the  filter scaling factor $n$  by measuring the fluxes  of 89 field stars in  both the $r'$ band  and S{\scriptsize II}
frames, assuming that field  stars show no H$\alpha+$[N{\scriptsize II}] emission and  hence provide the same level of continuum in  both filters. We measured an average flux  ratio of $1/n=29.05$. We
checked this value  by also computing the area under  both filter transmission curves, i.e.\  numerically integrating over the corresponding  filter transmission data. We used Simpson's  rule for that
purpose and  computed a similar ratio  of transmissivity between the  two filters of $1/n\,\approx29.03$,  confirming our measurements. We  checked the resulting H$\alpha+$[N{\scriptsize  II}] maps by
investigating the outer parts of early-type galaxies in our frames, not expected to show any H$\alpha+$[N{\scriptsize II}] emission.

We then converted countrates to physical fluxes using the following equation

\begin{equation}
\label{halphacalibration2}
{F_{{\rm{H}}\alpha+[{\rm{N {\scriptsize  II}]}},{\rm{obs.}}}} = \kappa \frac{{{c_{{\rm{H}}\alpha+[{\rm{N {\scriptsize  II}]}} {\rm{,corr.}}}}}}{{{R_{{\rm{SII}}}}({\rm{H}}\alpha )}}
\end{equation}

where  ${{c_{{\rm{H}}\alpha+[{\rm{N  {\scriptsize II}]}}  {\rm{,corr.}}}}}$  are  the extinction  corrected  net  H$\alpha+$[N{\scriptsize II}]  countrates,  $\kappa$  the photometric  zeropoint,  and
${{R_{{\rm{SII}}}}({\rm{H}}\alpha )}$  the sensitivity of the  narrowband filter at  the redshifted H$\alpha$  line. For bright H$\alpha$  sources, the contamination  of the continuum in  the brodband
filter by H$\alpha$ cannot be neglected. We corrected for this contamination using the following relation:

\begin{equation}
\label{halphalinecorrection}
{F_{{\rm{H}}\alpha+[{\rm{N {\scriptsize  II}]}},{\rm{0}}}} = {F_{{\rm{H}}\alpha+[{\rm{N {\scriptsize  II}]}} ,{\rm{obs.}}}}\left( {1 + \frac{{\int {{R_{{\rm{SII}}}}(\lambda )d\lambda } }}{{\int {{R_{r'}}(\lambda )d\lambda } }}} \right)
\end{equation}

where $F_{{\rm{H}}\alpha+[{\rm{N {\scriptsize  II}]}} {\rm{,obs.}}}$ is the observed flux and ${F_{{\rm{H}}\alpha+[{\rm{N {\scriptsize  II}]}} {\rm{,0}}}}$ the corrected one. 

\subsection{Flux measurements}
We measured  total galaxy magnitudes in  $g'$ and $r'$ as  well as H$\alpha+$[N{\scriptsize II}]  fluxes utilizing the Aperture  Photometry Tool\footnote{\url{http://www.aperturephotometry.org}} (APT)
package (\citealt{laher12}). Aperture radii were defined so  that the integrated, sky-subtracted galaxy flux reached a flat plateau in the outskirts of  the aperture curve of growth. Any remaining sky
level was  corrected by  subtracting the median  sky counts  measured within an  annulus around  each aperture. Interlopers  within the galaxy  aperture were  masked by hand.  To account  for Galactic
extinction we  used the  extinction maps from  \citet{schlaflyfinkbeiner11}. For  $g'$ and $r'$  we applied  the corresponding  extinction values as  listed in  NED. To compute  the extinction  in the
[S{\scriptsize II}] filter, we considered the extinction law of \citet{cardelli89} which can be paramterized as
\begin{equation}
\label{halphaextinction}
A_{\rm{X}} = \alpha_{\rm X} \cdot {R_V} \cdot E_{B - V}\\
\end{equation}

where $R_{V}=3.1$ is the extinction parameter,  $E_{B - V}$ the reddening from \citet{schlaflyfinkbeiner11}, and $\alpha_{\rm[SII]}=0.790$. 

\subsubsection{GALEX NUV and FUV measurements}
To complement our observations with fluxes  in the ultraviolet, we measured near-UV (NUV; $\lambda_{\rm eff}=2315.7$\AA) and far-UV (FUV;  $\lambda_{\rm eff}=1538.6$\AA) magnitudes from the background
subtracted intensity  maps found in the  GALEX GR6/7 Data  Release\footnote{\url{http://galex.stsci.edu/GR6/}} with the  same procedure as described  above. Except for  HCG26, NUV and FUV  images were
available for all groups in our  sample. With a spatial scale of 1.5 arcsec pixel$^{-1}$ and a spatial  resolution (FWHM) of 4.2 arcsec ($\sim 2.5$ kpc) in FUV and 5.3  arcsec ($\sim 3.2$ kpc) in NUV,
the UV images are coarse compared to our optical measurements. We computed NUV  and FUV magnitudes considering GALEX zeropoints $Z_{P,\rm FUV}=18.82$ and $Z_{P,\rm NUV}=20.08$. Galactic extinction was
estimated as above with $\alpha_{\rm NUV}=2.808$ and $\alpha_{\rm FUV}=2.617$. We considered one extinction value for each group. Table \ref{galextable} summarizes the utilized GALEX data.

\begin{table}
\begin{center}
\caption{GALEX NUV and FUV data.}
\begin{tabular}{ccc}
\hline                                                                                                                                                                                          
   group   &   survey$\,^{a}$    &     exposure time         \\      
\hline                                                             
  HCG04    &      AIS            &        \phantom{1}107s    \\                    
  HCG24    &      MIS            &                  2041s    \\                    
  HCG25    &      MIS            &                  3375s    \\                    
  HCG26    &      $\cdots$       &            $\cdots$       \\                    
  HCG89    &      MIS            &                  2427s    \\                    
  HCG91    &      AIS            &        \phantom{1}191s    \\                    
  HCG96    &      NGS            &                  1643s    \\                      
\hline
\end{tabular}
\label{galextable}
\end{center}
{\bf Notes:} $^{a}\,$AIS -- All Sky Imaging Survey.\\ MIS -- Medium Imaging Survey.\\ NGS -- Nearby Galaxies Survey.
\end{table}

\subsubsection{Error bars\label{errorbars}}
We estimated error bars within the APT photometry package  based on equation 3 shown in \citet{laher12}. To do so, we first computed the effective gain  $\mathcal G = N \times t_{\rm exp}\times G$ for
each image, where  $N$ is the number of  averaged frames, $t_{\rm exp}$ the  average exposure time, and $G$  the gain of the raw  images. For all of our  measurements in $g'$ and $r'$,  the error bars
estimated this way are much smaller than the  accuracy of the derived photometric zeropoints. Hence, we consider the uncertainty in the photometric calibration, 0.08  mag in $r'$ and 0.12 mag in $g'$,
as error estimate for all of our integrated $g'$ and $r'$ measurements (see Sect.\ \ref{photcal}). Consequently, we estimate an error of 0.14 mag for $g'$$-$$r'$ colours.

\subsection{Detection limits}
We estimated detection limits  for the utilized $g'$, $r'$, NUV  and FUV images and for the  created H$\alpha+$[N{\scriptsize II}] maps. To do so,  we sampled the scatter in the  sky background of the
corresponding frames with  {\sevensize IRAF imexam} and considered a  value of $3\sigma$ as detection  limit. Applying the appropriate zero-points  and spatial scales we converted these  limits to mag
arcsec$^{-2}$ for  $g'$, $r'$,  NUV and FUV  and to $10^{-16}$  erg s$^{-1}$  cm$^{-2}$ arcsec$^{-2}$ for  the H$\alpha+$[N{\scriptsize  II}] maps. The  resulting detection limits  are shown  in Table
\ref{completeness3sigma}.

\begin{table}
\begin{center}
\caption{Detection limits.}
\begin{tabular}{cccccc}
\hline                                                                                                                                                                                          
group   &  $g'\,^{a}$      &  $r'\,^{a}$       &     NUV$\,^{a}$     &        FUV$\,^{a}$  &     H$\alpha+$[N{\scriptsize  II}]$^{b}$   \\       
\hline                                                             
HCG04   &      24.68      &      24.30       &       25.48         &        25.45        &                    0.61                    \\           
HCG24   &      24.88      &      24.10       &       26.71         &        26.55        &                    0.99                    \\           
HCG25   &      25.04      &      24.31       &       27.15         &        27.04        &                    4.69                    \\           
HCG26   &      24.99      &      24.25       &      $\cdots$       &      $\cdots$       &                    1.55                    \\           
HCG89   &      23.99      &      23.82       &       26.99         &        26.94        &                    1.10                    \\           
HCG91   &      23.88      &      24.08       &       25.64         &        25.81        &                    1.86                    \\           
HCG96   &      24.42      &      23.91       &       26.80         &        26.91        &                    0.83                    \\             
\hline
\end{tabular}
\label{completeness3sigma}
\end{center}
{\bf Notes:} $^{a}\,$$g'$, $r'$, NUV and FUV limits are given in mag arcsec$^{-2}$.\\
             $^{b}\,$H$\alpha+$[N{\scriptsize  II}] limits are given in $10^{-16}$ erg s$^{-1}$ cm$^{-2}$ arcsec$^{-2}$.
\end{table}

\section{Data analysis and results}
\begin{figure*}
\begin{center}
\begin{sideways}
\begin{minipage}{21cm}
\includegraphics[width=\textwidth]{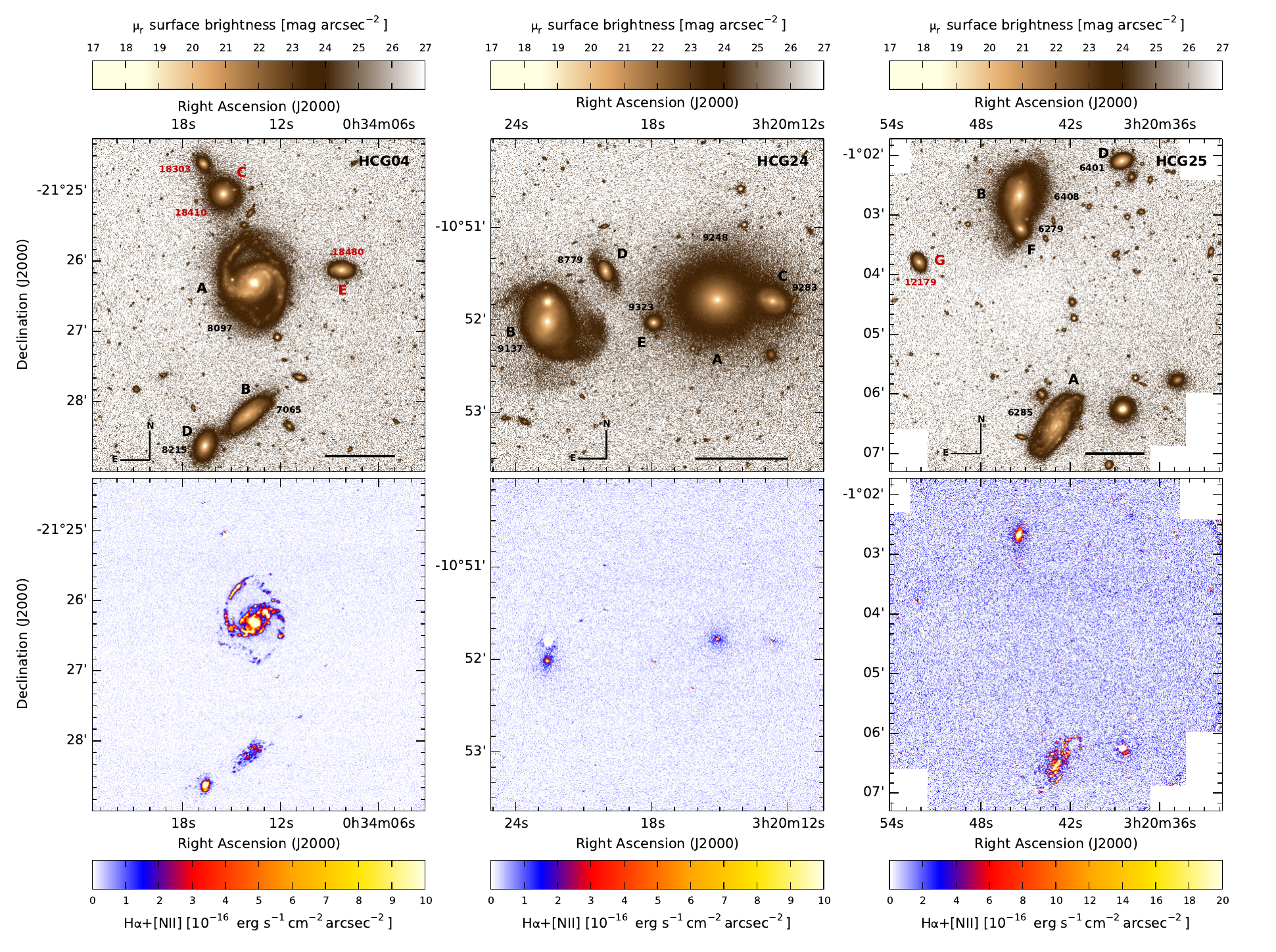}
\caption{\label{gallery1}Observed Hickson Compact Groups in the  $r'$ band (top panels) and the corresponding continuum corrected H$\alpha+$[N{\scriptsize II}]  maps (lower panels). The colorscale for
the $r'$ band  frames shows  the proper  surface brightness  in mag  arcsec$^{-2}$. Radial  velocities of group  members, taken  from SIMBAD,  are also  shown. Galaxies  with discordant  redshifts are
highlighted in red. The  horizontal black lines indicate 1 arcmin. H$\alpha+$[N{\scriptsize II}]  fluxes are given in $10^{-16}$ erg s$^{-1}$ cm$^{-2}$  arcsec$^{-2}$.}
\end{minipage}
\end{sideways}
\end{center}
\end{figure*}

\begin{figure*}
\begin{sideways}
\begin{minipage}{21cm}
\begin{center}
\includegraphics[width=\textwidth]{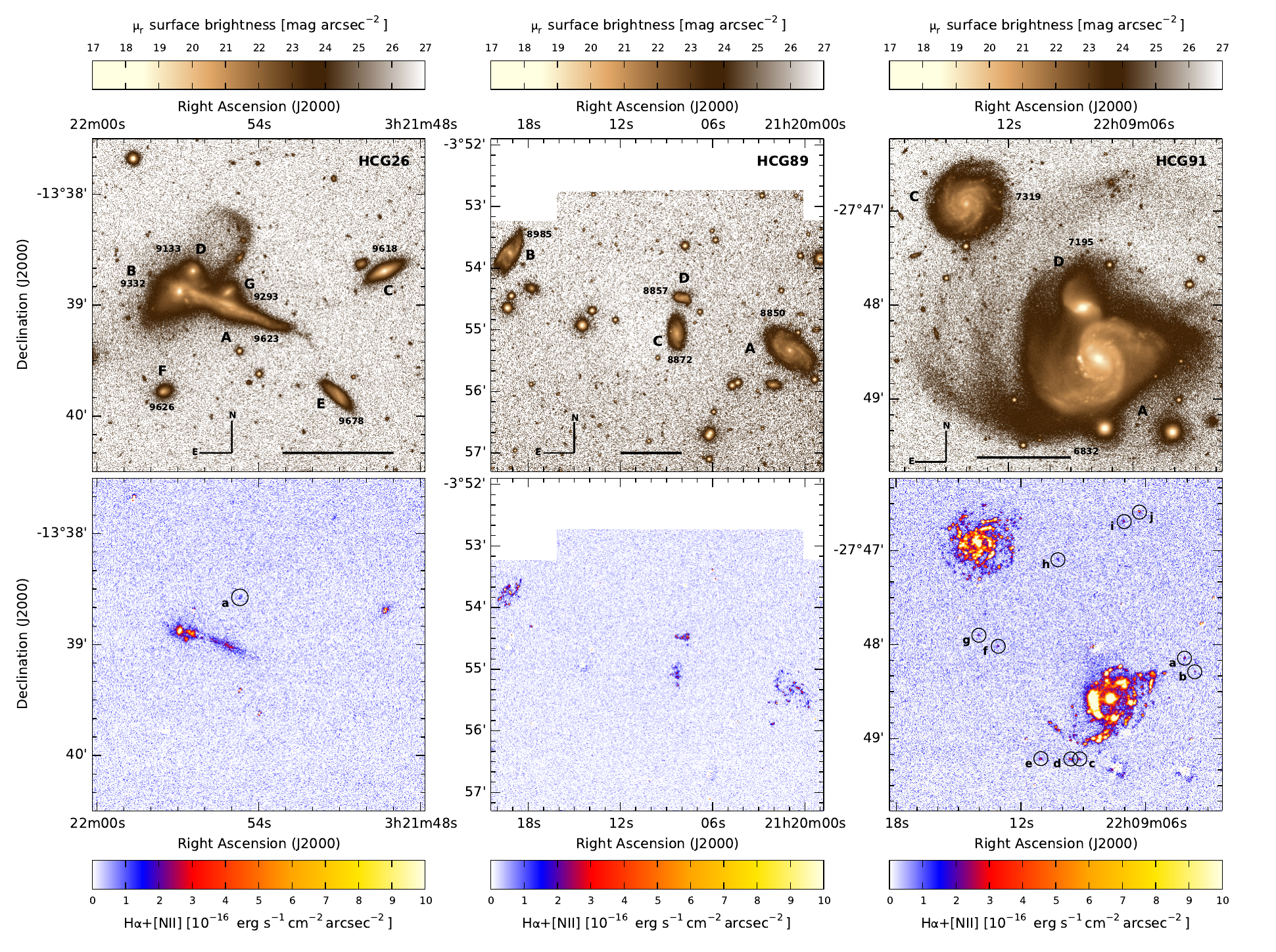}
\caption{\label{gallery2} Same as Fig.\ \ref{gallery1} but for HCG26, HCG89, and HCG91. Star-forming regions associated with tidal tails are marked with circles. See Sect.\ \ref{tdgsection}.}
\end{center}
\end{minipage}
\end{sideways}
\end{figure*}

\begin{figure}
\begin{center}
\includegraphics[width=0.7841\columnwidth]{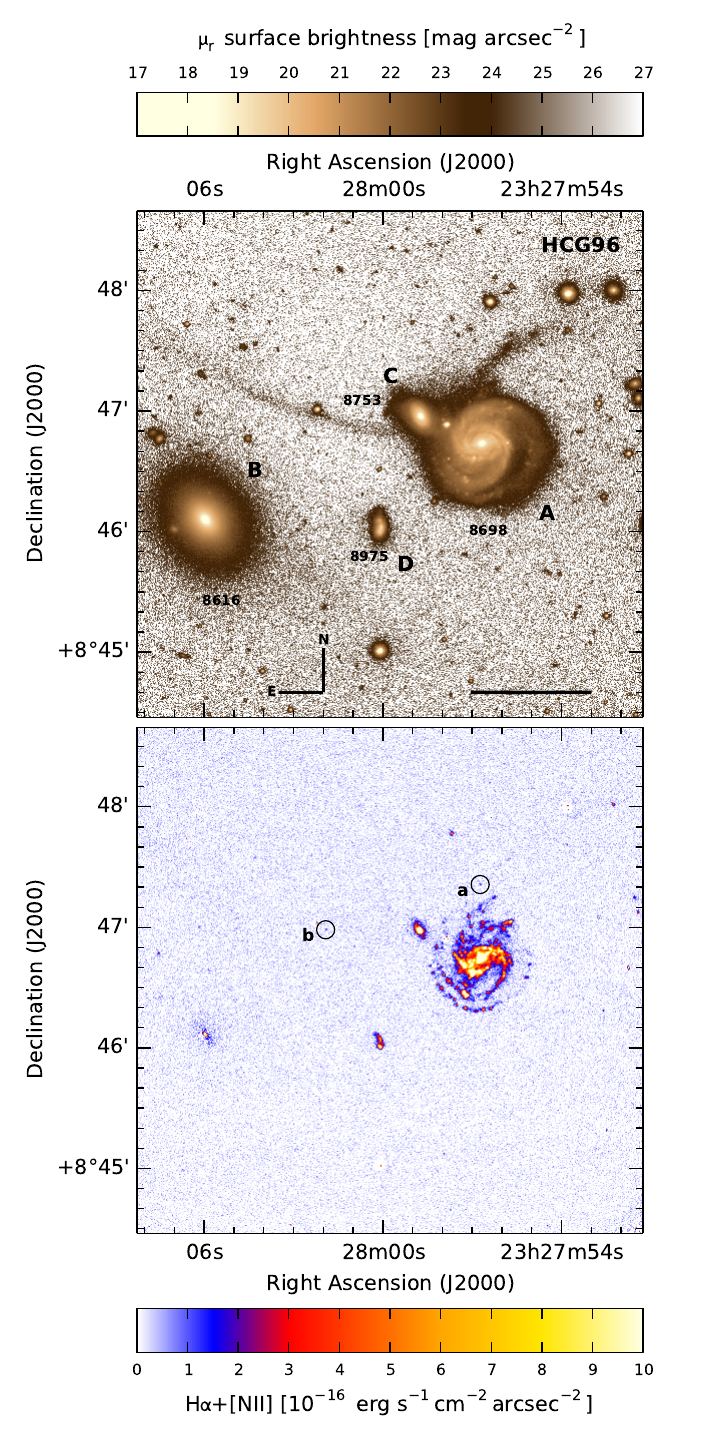}
\end{center}
\caption{\label{gallery3} Same as Fig.\ \ref{gallery1} but for HCG96. Star-forming regions associated with tidal tails are marked with circles. See Sect.\ \ref{tdgsection}.}
\end{figure}

Table  \ref{photometrictable} shows  the integrated,  total $g'$,  $r'$, NUV  and  FUV magnitudes  as well  as $g'$$-$$r'$  and NUV$-r'$  colors  for all  bright galaxies  in our  HCG sample.  Figures
\ref{gallery1}, \ref{gallery2} and \ref{gallery3}  show the observed $r'$ band frames in mag  arcsec$^{-2}$ and the corresponding, continuum corrected H$\alpha+$[N{\scriptsize  II}] maps in $10^{-16}$
erg s$^{-1}$ cm$^{-2}$ arcsec$^{-2}$. We note that bright stars or the bright cores of  some ellipticals produce artefacts in the created H$\alpha+$[N{\scriptsize II}] maps possibly due to a different
PSF in different passbands. When measured, these artefacts don't show any noteworthy H$\alpha+$[N{\scriptsize II}] emission, however.


\begin{table*}
\begin{center}
\caption{Photometric measurements, radial velocities, and stellar masses of bright HCG galaxies.}
\begin{tabular}{cccccccccccc}
\hline                                                                                                                                                                                          
\multirow{2}{*}{group}              &  \multirow{2}{*}{galaxy} &    \multirow{2}{*}{$\alpha_{2000}$} &  \multirow{2}{*}{$\delta_{2000}$}  &   $g'\,^{a}$         &    $r'\,^{a}$         &  $g'$$-$$r'\,^{a}$   &  NUV               &  NUV-$r'$            &    $v_{r}\,^b$       & $\log\,$M$_{\star}$    &  \multirow{2}{*}{type$^{d}$}         \\    
                                    &                          &                                     &                                    &   [mag]             &    [mag]             &  [mag]             &  [mag]             &  [mag]              &    [km s$^{-1}$]     & [M$_{\odot}$]          &                                      \\    
\hline                                                             
\multicolumn{1}{c}{\multirow{5}{*}{HCG04}}    &             A\dotfill              &                  00 34 13.64                  &                   $-$21 26 19.0                &         13.26                        &         12.82                        &           0.44                           &    15.13$\,\pm\,$0.01             &        2.31$\,\pm\,$0.08      &              8097                    &             11.14$\,\pm\,$0.06    &                   Sc  \dotfill (1)                                   \\    
                                              &             B\dotfill              &                  00 34 14.07                  &                   $-$21 28 12.0                &         15.15                        &         14.64                        &           0.52                           &    17.22$\,\pm\,$0.04             &        2.59$\,\pm\,$0.09      &              7065                    &             10.40$\,\pm\,$0.07    &                   Sb  \dotfill (2)                                   \\    
                                              &             C\dotfill              &                  00 34 15.50                  &                   $-$21 25 03.0                &         15.38                        &         14.61                        &           0.77                           &    21.02$\,\pm\,$0.34             &        6.41$\,\pm\,$0.35      &             18410$^*$$^{c}$          &             11.41$\,\pm\,$0.03    &                   E2  \dotfill (1)                                   \\    
                                              &             D\dotfill              &                  00 34 16.65                  &                   $-$21 28 38.0                &         15.51                        &         14.89                        &           0.62                           &    17.67$\,\pm\,$0.05             &        2.79$\,\pm\,$0.09      &              8215                    &             10.44$\,\pm\,$0.04    &                   E4  \dotfill (1)                                   \\    
                                              &             E\dotfill              &                  00 34 08.27                  &                   $-$21 26 08.1                &         15.93                        &         15.31                        &           0.61                           &    18.44$\,\pm\,$0.06             &        3.13$\,\pm\,$0.10      &             18480$^*$                &             11.04$\,\pm\,$0.06    &                   Sab \dotfill (2)                                   \\    
\hline                                                                                                                                                                                                                                                                                                                                                                                                                                                                                    
\multicolumn{1}{c}{\multirow{5}{*}{HCG24}}    &             A\dotfill              &                  03 20 15.14                  &                   $-$10 51 46.8                &         14.57                        &         13.73                        &           0.84                           &    20.07$\,\pm\,$0.08             &        6.34$\,\pm\,$0.11      &              9248                    &             11.19$\,\pm\,$0.07    &                   S0  \dotfill (2)                                   \\    
                                              &             B\dotfill              &                  03 20 22.61                  &                   $-$10 52 00.8                &         14.61                        &         13.93                        &           0.68                           &    20.74$\,\pm\,$0.13             &        6.80$\,\pm\,$0.15      &              9137                    &             11.04$\,\pm\,$0.06    &                   SBa \dotfill (1)                                   \\    
                                              &             C\dotfill              &                  03 20 12.77                  &                   $-$10 51 47.8                &         16.05                        &         15.04                        &           1.01                           &    20.88$\,\pm\,$0.14             &        5.84$\,\pm\,$0.16      &              9283                    &             10.71$\,\pm\,$0.23    &                   SB0 \dotfill (2)                                   \\    
                                              &             D\dotfill              &                  03 20 20.10                  &                   $-$10 51 28.8                &         16.53                        &         16.02                        &           0.51                           &    21.27$\,\pm\,$0.21             &        5.25$\,\pm\,$0.23      &              8779                    &   \phantom{1}9.98$\,\pm\,$0.06    &                   S0a \dotfill (2)                                   \\    
                                              &             E\dotfill              &                  03 20 17.93                  &                   $-$10 52 01.8                &         17.26                        &         16.50                        &           0.76                           &    21.82$\,\pm\,$0.36             &        5.32$\,\pm\,$0.37      &              9323                    &             10.04$\,\pm\,$0.06    &                   E   \dotfill (2)                                   \\    
\hline                                                                                                                                                                                                                                                                                                                                                                                                                                                       
\multicolumn{1}{c}{\multirow{5}{*}{HCG25}}    &             A\dotfill              &                  03 20 42.95                  &                   $-$01 06 30.7                &         14.03                        &         13.62                        &           0.41                           &    15.76$\,\pm\,$0.01             &        2.14$\,\pm\,$0.08      &              6285                    &             10.50$\,\pm\,$0.03    &                   SBc \dotfill (1)                                   \\    
                                              &             B\dotfill              &                  03 20 45.42                  &                   $-$01 02 40.8                &         14.29                        &         13.34                        &           0.96                           &    18.78$\,\pm\,$0.05             &        5.44$\,\pm\,$0.09      &              6408                    &             11.12$\,\pm\,$0.09    &                   SBa \dotfill (1)                                   \\    
                                              &             D\dotfill              &                  03 20 38.81                  &                   $-$01 02 07.4                &         15.44                        &         14.80                        &           0.64                           &    19.92$\,\pm\,$0.07             &        5.12$\,\pm\,$0.11      &              6401                    &             10.25$\,\pm\,$0.04    &                   S0  \dotfill (1)                                   \\    
                                              &             F\dotfill              &                  03 20 45.35                  &                   $-$01 03 13.8                &         15.77                        &         15.17                        &           0.60                           &    20.93$\,\pm\,$0.07             &        5.76$\,\pm\,$0.11      &              6279                    &             10.15$\,\pm\,$0.08    &                   S0  \dotfill (1)                                   \\    
                                              &             G\dotfill              &                  03 20 52.10                  &                   $-$01 03 47.3                &         16.31                        &         15.51                        &           0.79                           &    21.38$\,\pm\,$0.13             &        5.86$\,\pm\,$0.15      &             12179$^*$                &             10.68$\,\pm\,$0.06    &                   S0  \dotfill (2)                                   \\    
\hline                                                                                                                                                                                                                                                                                                                                                                                                                                                            
\multicolumn{1}{c}{\multirow{7}{*}{HCG26}}    &             A\dotfill              &                  03 21 55.30                  &                   $-$13 39 03.1                &         15.26                        &         14.69                        &           0.58                           &         $\cdots$                  &            $\cdots$           &              9678                    &             10.60$\,\pm\,$0.01    &                   Scd \dotfill (1)                                   \\    
                                              &             B\dotfill              &                  03 21 57.15                  &                   $-$13 38 54.3                &         16.07                        &         15.16                        &           0.91                           &         $\cdots$                  &            $\cdots$           &              9332                    &             10.79$\,\pm\,$0.02    &                   E0  \dotfill (1)                                   \\    
                                              &             C\dotfill              &                  03 21 49.42                  &                   $-$13 38 41.7                &         16.46                        &         15.70                        &           0.76                           &         $\cdots$                  &            $\cdots$           &              9618                    &             10.39$\,\pm\,$0.02    &                   S0  \dotfill (2)                                   \\    
                                              &             D\dotfill              &                  03 21 56.50                  &                   $-$13 38 42.0                &         16.90                        &         16.29                        &           0.61                           &         $\cdots$                  &            $\cdots$           &              9133                    &             10.01$\,\pm\,$0.01    &                   cI  \dotfill (1)                                   \\    
                                              &             E\dotfill              &                  03 21 51.15                  &                   $-$13 39 48.5                &         17.26                        &         16.82                        &           0.44                           &         $\cdots$                  &            $\cdots$           &              9623                    &   \phantom{1}9.60$\,\pm\,$0.01    &                   Im  \dotfill (1)                                   \\    
                                              &             F\dotfill              &                  03 21 57.65                  &                   $-$13 39 46.7                &         17.94                        &         17.53                        &           0.40                           &         $\cdots$                  &            $\cdots$           &              9626                    &   \phantom{1}9.27$\,\pm\,$0.01    &                   cI  \dotfill (2)                                   \\    
                                              &             G\dotfill              &                  03 21 55.23                  &                   $-$13 38 55.5                &         17.62                        &         16.88                        &           0.74                           &         $\cdots$                  &            $\cdots$           &              9293                    &   \phantom{1}9.91$\,\pm\,$0.01    &                   S0  \dotfill (2)                                   \\    
\hline                                                                                                                                                                                                                                                                                                                                                                                                                                                              
\multicolumn{1}{c}{\multirow{4}{*}{HCG89}}    &             A\dotfill              &                  21 20 01.03                  &                   $-$03 55 19.6                &         14.70                        &         14.28                        &           0.43                           &    16.74$\,\pm\,$0.01             &        2.46$\,\pm\,$0.08      &              8850                    &            10.58$\,\pm\,$0.04     &                   Sc  \dotfill (2)                                   \\    
                                              &             B\dotfill              &                  21 20 19.12                  &                   $-$03 53 45.6                &         15.44                        &         14.96                        &           0.47                           &    17.21$\,\pm\,$0.01             &        2.25$\,\pm\,$0.08      &              8985                    &            10.29$\,\pm\,$0.05     &                   SBc \dotfill (1)                                   \\    
                                              &             C\dotfill              &                  21 20 08.31                  &                   $-$03 55 03.6                &         15.78                        &         15.38                        &           0.40                           &    17.83$\,\pm\,$0.02             &        2.45$\,\pm\,$0.08      &              8872                    &            10.14$\,\pm\,$0.07     &                   Scd \dotfill (2)                                   \\    
                                              &             D\dotfill              &                  21 20 08.00                  &                   $-$03 54 30.0                &         16.41                        &         16.23                        &           0.18                           &    17.74$\,\pm\,$0.02             &        1.51$\,\pm\,$0.08      &              8857                    &  \phantom{1}9.49$\,\pm\,$0.01     &                   Sm  \dotfill (2)                                   \\    
\hline                                                                                                                                                                                                                                                                                                                                                                                                                                             
\multicolumn{1}{c}{\multirow{3}{*}{HCG91}}    &             A\dotfill              &                  22 09 07.69                  &                   $-$27 48 34.0                &         12.69                        &         12.05                        &           0.64                           &    14.78$\,\pm\,$0.01             &        2.72$\,\pm\,$0.08      &              6832                    &            11.50$\,\pm\,$0.03     &                   SBc \dotfill (1)                                   \\    
                                              &             C\dotfill              &                  22 09 14.02                  &                   $-$27 46 56.0                &         14.46                        &         13.99                        &           0.47                           &    16.18$\,\pm\,$0.02             &        2.19$\,\pm\,$0.08      &              7319                    &            10.52$\,\pm\,$0.03     &                   Sc  \dotfill (2)                                   \\    
                                              &             D\dotfill              &                  22 09 08.20                  &                   $-$27 48 01.0                &         14.54                        &         13.77                        &           0.77                           &    19.04$\,\pm\,$0.07             &        5.27$\,\pm\,$0.10      &              7195                    &            10.93$\,\pm\,$0.01     &                   SB0 \dotfill (1)                                   \\    
\hline                                                                                                                                                                                                                                                                                                                                                                                                                                                    
\multicolumn{1}{c}{\multirow{4}{*}{HCG96}}    &             A\dotfill              &                  23 27 56.70                  &         \phantom{$-$}08 46 44.4                &         13.25                        &         12.78                        &           0.48                           &    15.45$\,\pm\,$0.01             &        2.67$\,\pm\,$0.08      &              8698                    &            11.34$\,\pm\,$0.14    &                   Sc  \dotfill (1)                                   \\    
                                              &             B\dotfill              &                  23 28 05.95                  &         \phantom{$-$}08 46 07.4                &         14.12                        &         13.23                        &           0.89                           &    19.42$\,\pm\,$0.05             &        6.19$\,\pm\,$0.09      &              8616                    &            11.40$\,\pm\,$0.06    &                   E2  \dotfill (1)                                   \\    
                                              &             C\dotfill              &                  23 27 58.80                  &         \phantom{$-$}08 46 58.4                &         15.15                        &         14.51                        &           0.64                           &    18.54$\,\pm\,$0.02             &        4.03$\,\pm\,$0.08      &              8753                    &            10.78$\,\pm\,$0.09    &                   Sa  \dotfill (1)                                   \\    
                                              &             D\dotfill              &                  23 28 00.20                  &         \phantom{$-$}08 46 02.0                &         16.41                        &         16.03                        &           0.38                           &    18.05$\,\pm\,$0.02             &        2.02$\,\pm\,$0.08      &              8975                    &  \phantom{1}9.78$\,\pm\,$0.01    &                   Im  \dotfill (1)                                   \\      
\hline
\end{tabular}
\label{photometrictable}
\end{center}
{\bf Notes:} $^{a}\,$We consider error bars of 0.08 mag in $r'$, 0.12 mag in $g'$, and 0.14 mag in $g'$$-$$r'$ (see Sect.\ \ref{errorbars} for details.)
             $^{b}\,$Radial velocities taken from  \citet{hickson92}. Asterisks show nonmember galaxies originally classified as group members by \citet{hickson82}. 
             $^{c}\,$For HCG04C we list the radial velocity from SIMBAD, measured much more recently and showing a severe mismatch compared to the radial velocity given in \citet{hickson92} (8863 km s$^{-1}$).
             $^{d}\,$(1): morphology taken from \citet{claudia94} --- (2): morphology taken from SIMBAD.
\end{table*}

\subsection{Compact group galaxies in the NUV-$r'$ plane}
\begin{figure}
\begin{center}
\includegraphics[width=220pt]{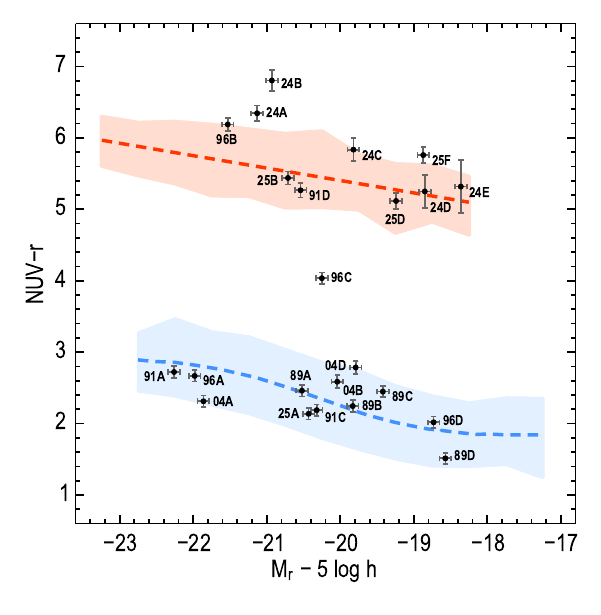}
\end{center}
\caption{\label{greenvalley} NUV$-r'$ vs. $M_{r'}$ colour-magnitude  diagram for all observed compact group member galaxies.  The dashed lines indicate the red and blue  sequences for SDSS galaxies in
the local universe derived by \citet{wyder07}. Compact group  members from our sample clearly fall on these sequences except for one outlier, HCG96C,  located inside the green valley. The dashed areas
show $1\sigma$ confidence intervals. The interlopers falsely classified as group members by \citet{hickson82} are not shown in the plot.}
\end{figure}

To distinguish between  actively star-forming galaxies and passive quiescent  galaxies, we constructed a NUV$-r'$ vs.  $M_{r'}$ colour magnitude diagram (CMD). Figure  \ref{greenvalley} shows that our
compact  group galaxies  are clearly  separated in  the NUV$-r'$  vs.\ $M_{r'}$  plane, following  the well-defined  blue and  red sequences  (dashed lines)  for SDSS  galaxies in  the local  universe
\citep{wyder07}. We  only find one  galaxy, HCG96C,  located in the  green valley. From  a total of  23 bright  HCG member galaxies  in our sample  (for which we  have NUV  data), we find  12 galaxies
($\sim52$\%) located in the blue sequence while we find 10 ($\sim44$\%) located in the red sequence  and one outlier ($\sim4$\%) in the green valley. For our further analysis we only consider galaxies
in the blue sequence as star-forming and unless  otherwise stated compare only these star-formation rates with the literature. Since we do not have any  UV data for HCG26, we use the $g'$$-$$r'$ color
criterion proposed by \citet{bell03} for this system to distinguish star-forming and passive galaxies.

\subsection{Star formation rates\label{starformationrates}}
To properly convert  the measured H$\alpha+$[N{\scriptsize II}]  narrowband fluxes into star-formation  rates, two corrections have  to be taken into account.  First, the narrowband fluxes  have to be
corrected for the contamination by  the nitrogen [N{\scriptsize II}]$\lambda\lambda6548,6583$ lines. Second, the remaining  H$\alpha$ fluxes have to be corrected for extinction  due to dust within the
observed galaxies. To apply these corrections, we utilize the following procedure.

Correcting for  the contamination by the  nitrogen [N{\scriptsize II}]$\lambda\lambda6548,6583$ lines  is challenging since the  correction factor H$\alpha$/(H$\alpha+$[N{\scriptsize II}])  can show a
considerable  scatter  when  measuring different  HII  regions  within  one  galaxy.  Based  on the  integrated  [N{\scriptsize  II}]/H$\alpha$  measurements  of  58  galaxies from  the  SINGS  survey
(\citealt{kennicutt03,kennicutt09}), we consider an  average H$\alpha$/(H$\alpha+$[N{\scriptsize II}]) ratio of 0.74 $\pm$ 0.13  for the star-forming galaxies in our sample. We  also checked if any of
our galaxies is found in SDSS with spectroscopic measurements  to verify this correction factor. We only found SDSS spectra for galaxies HCG25A and B and  found that HCG25B hosts an AGN. For HCG25A we
measured a [N{\scriptsize II}]/H$\alpha$ flux ratio of 0.69 which is in perfect agreement with the average value from the SINGS sample, proving the reliability of the applied correction factor.

Second, we account  for H$\alpha$ extinction due to dust  within the observed galaxies. This  correction is difficult since the dust  fraction can vary strongly in galaxies  of different morphological
type and can even show a considerable  scatter within a single galaxy. Recent studies have revealed a clear trend between  H$\alpha$ extinction and stellar mass (\citealt{garnbest10,kashino}), arguing
that the knowledge of stellar mass is sufficient  to model the extinction of a galaxy in a statistical sense, i.e.\ once the mass dependence  is applied, the accuracy of the extinction estimate cannot
be improved significantly by a more complex model. We use the relation of \citet{garnbest10}

\begin{equation}
{A_{{\rm{H}}\alpha }} = 0.91 + 0.77X + 0.11{X^2} - 0.09{X^3}
\end{equation}

to  account for  a  stellar mass  dependent  H$\alpha$ dust  extinction  correction, where  $X  =  \log ({M_  *  }/{10^{10}}{{\rm{M}}_ \odot  })$.  The authors  state  an uncertainty  of  0.28 mag  in
${A_{{\rm{H}}\alpha }}$. The final corrected H$\alpha$ fluxes were subsequently converted to H$\alpha$ luminosities via:

\begin{equation}
\label{starformation}
{L_{{\rm{H}}\alpha }} = 4{r^2}\pi  {F_{{\rm{H}}\alpha }}
\end{equation}

where $r$  is the corresponding luminosity  distance in cm and  $F_{{\rm H}\alpha}$ the integrated  H$\alpha$ flux of the  whole galaxy. We computed  star-formation rates in  M$_\odot$ year$^{-1}$ by
applying the calibration for H$\alpha$ fluxes as presented in \citet{kennicutt12}:

\begin{equation}
\label{starformation}
\log {\dot M_*}[{{\rm{M}}_ \odot\; }{\rm{yea}}{{\rm{r}}^{ - 1}}] = \log {L_{{\rm H}\alpha}}[{\rm{erg\; }}{{\rm{s}}^{ - 1}}] -  41.27
\end{equation}

We estimated  error bars considering  the uncertainties in  both the H$\alpha$/(H$\alpha+$[N{\scriptsize  II}]) correction  and the dust  extinction since these  error sources are  considerably larger
compared to the uncertainties in the flux measurements.

\subsubsection{Correcting for AGN contamination\label{agn}}
We checked the  literature if any of our galaxies  host an AGN to correct for  any contamination in the measured emission  line fluxes caused by non photo-ionization.  \citet{martinez10} carried out a
spectroscopic survey to identify AGNs  in HCG galaxies, revealing that some of our  objects indeed host an AGN. To get a  more complete census of AGN contamination in our  sample, we also searched the
whole catalog of quasars and active nuclei  from \citet{veroncetty} but could not find any additional AGNs. Based on our analysis of  its SDSS spectrum, we measure $\log$([N{\scriptsize II}]/H$\alpha)
=  0.26$ for  the elliptical  HCG25B. Comparing  this  value with  the diagnostic  diagram  for emission  line galaxies  in \citet{brinchmann04},  HCG25B  can definitely  be considered  as AGN.  Table
\ref{starformationtable} lists the activity type from \citet{martinez10} for the star-forming galaxies in  our sample. Galaxies are classified as active galactic nuclei (AGN), transition objects (TO),
and star-forming nuclei (SFN),  where TOs are emission-line galaxies with line ratios  intermediate between SFNs and AGNs. For galaxies hosting  an AGN or TO we exclude the flux  of the very center of
each galaxy. Since the physical scale of an  AGN is below our spatial resolution, we consider a central aperture matching the seeing of our  observations, i.e.\ $\sim0.8$ arcsec and subtract this flux
from our integrated measurements.

\subsubsection{Comparison with FUV}
\begin{figure}
\begin{center}
\includegraphics[width=220pt]{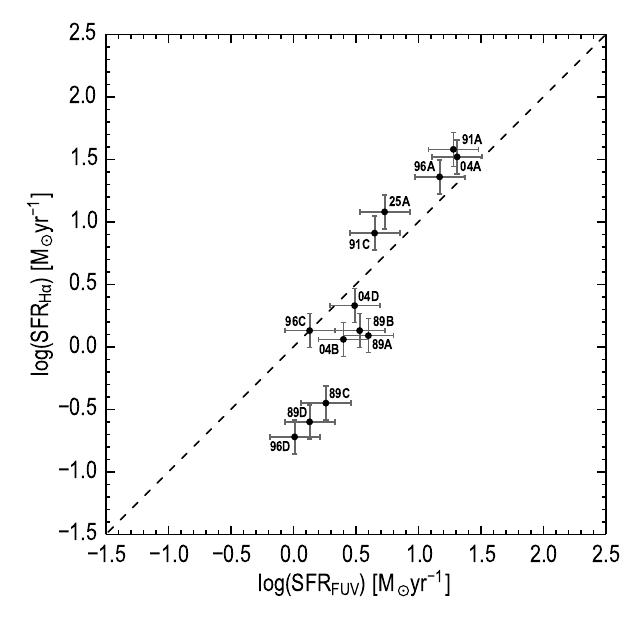}
\end{center}
\caption{\label{SFRcomparison} Comparison between FUV and H$\alpha$ star-formation  rates for all compact group galaxies on the NUV$-r'$ blue sequence. There  is a good agreement between H$\alpha$ and
FUV star-formation rates for the brightest star-forming galaxies in our sample, while fainter galaxies show higher star-formation rates in FUV compared to H$\alpha$.}
\end{figure}

For comparison, we also derived  star-formation rates from our FUV measurements. The main disadvantage of  FUV as a tracer for star-formation is the strong FUV  extinction due to interstellar dust. We
correct for this attenuation considering FUV$-$NUV colors. According to \citet{salim07}, FUV extinction $A_{\rm FUV}$ can be parametrized as

\begin{equation}
\label{starformation}
{A_{{\rm{FUV}}}} = 2.99\cdot({\rm{FUV}} - {\rm{NUV)}} + 0.27
\end{equation}

for galaxies with NUV$-r'<4$ and $({\rm FUV}-{\rm NUV})<0.90$,  which is true for all of our blue sequence objects. The typical scatter in  $A_{{\rm{FUV}}}$ is 0.5 mag (Salim, private communication --
cf.\ Fig.\ 13 in  \citealt{salim07}), which we consider as the main  error source for our FUV SFR estimates. After  applying this dust extinction correction, we converted  FUV$_{\rm AB}$ magnitudes to
fluxes in erg s$^{-1}$ cm$^{-2}$ Hz$^{-1}$, then computed FUV luminosities in erg s$^{-1}$ Hz$^{-1}$, and finally considered the relation of \citet{salim07}

\begin{equation}
\label{starformation}
{\dot M_*}[{{\rm{M}}_ \odot\; }{\rm{yea}}{{\rm{r}}^{ - 1}}] = 1.08 \times 10^{-28} {L_{{\rm FUV}}}[{\rm{erg\; }}{{\rm{s}}^{ - 1}}{\rm Hz}^{-1}]
\end{equation}

assuming a Salperter IMF. Figure \ref{SFRcomparison} shows  the comparison of star-formation rates derived from integrated H$\alpha$ and FUV fluxes for all  compact group galaxies in the NUV$-r'$ blue
sequence while Table \ref{starformationtable} lists all the corresponding values. We get a good  agreement between H$\alpha$ and FUV star-formation rates for the brightest star-forming galaxies in our
sample, while fainter galaxies show higher star-formation rates in FUV compared to H$\alpha$.

\begin{table*}
\begin{center}
\caption{Star-formation rates of HCG member galaxies in the NUV-r blue sequence derived from H$\alpha$ and FUV fluxes.}
\begin{tabular}{cccccccccc}
\hline                                                                                                                                                                                          
\multirow{2}{*}{group}                              &  \multirow{2}{*}{galaxy}           &   $\log {F_{{\text{H}}\alpha  + {\text{NII}}}}$$^{a}$             &       $\log {L_{{\rm{H}}\alpha }}$$^{b}$                &  $\log$ SFR$_{{\rm H}\alpha}$$^{b}$             &    FUV                           &    $\log {L_{{\rm{FUV}}}}$$^{c}$             &    $\log$ SFR$_{{\rm FUV}}$$^{c}$             &  $\log$ SSFR$_{{\rm H}\alpha}$      &      \multirow{2}{*}{activity$^{d}$ }  \\     
                                                    &                                    &   [erg s$^{-1}$cm$^{-2}$]                                         &       [erg s$^{-1}$]                                    &  [M$_{\odot}$$\,$year$^{-1}$]                   &    [mag]                         &    [erg s$^{-1}$ Hz$^{-1}$]                  &    [M$_{\odot}$$\,$year$^{-1}$]               &  [Gyear$^{-1}$]                     &                                        \\                         
\hline                                                                                                                                                                                                                                                                                                                                                                                                                                                                                    
\multicolumn{1}{c}{\multirow{3}{*}{HCG04}}          &             A\dotfill              &                    $-$12.06                                       &                     42.79                               &           \phantom{$-$}1.52                     &      15.63$\,\pm\,$0.03          &             29.27                            &        \phantom{$-$}1.31                      &        $-$0.62$\,\pm\,$0.15         &         TO                             \\     
                                                    &             B\dotfill              &                    $-$13.29                                       &                     41.33                               &           \phantom{$-$}0.06                     &      17.64$\,\pm\,$0.07          &             28.36                            &        \phantom{$-$}0.40                      &        $-$1.34$\,\pm\,$0.15         &         TO                             \\     
                                                    &             D\dotfill              &                    $-$13.03                                       &                     41.60                               &           \phantom{$-$}0.33                     &      18.43$\,\pm\,$0.09          &             28.46                            &        \phantom{$-$}0.49                      &        $-$1.11$\,\pm\,$0.14         &         SFN                            \\     
\hline                                                                                                                                                                                                                                                                                                                                                                                                                                                                                    
\multicolumn{1}{c}{\multirow{2}{*}{HCG25}}          &             A\dotfill              &                    $-$12.05                                       &                     42.35                               &           \phantom{$-$}1.08                     &      16.16$\,\pm\,$0.01          &             28.70                            &       \phantom{$-$}0.73                       &       $-$0.42$\,\pm\,$0.14          &         \dots                          \\     
                                                    &             B\dotfill              &                    $-$12.50                                       &                     42.09                               &           \phantom{$-$}0.82                     &      20.43$\,\pm\,$0.13          &             27.74                            &                $-$0.23                        &        $-$1.30$\,\pm\,$0.16         &         AGN*                           \\     
\hline                                                                                                                                                                                                                                                                                                                                                                                                                                                                                    
\multicolumn{1}{c}{\multirow{1}{*}{HCG26}$^{e}$}    &             A\dotfill              &                    $-$13.33                                       &                     41.46                               &           \phantom{$-$}0.19                     &      $\cdots$                    &            $\cdots$                          &        \phantom{$-$}$\cdots$                  &        $-$1.41$\,\pm\,$0.14         &         \dots                          \\     
\hline                                                                                                                                                                                                                                                                                                                                                                                                                                                                               
\multicolumn{1}{c}{\multirow{4}{*}{HCG89}}          &             A\dotfill              &                    $-$13.36                                       &                     41.36                               &           \phantom{$-$}0.09                     &      17.09$\,\pm\,$0.01          &             28.57                            &        \phantom{$-$}0.60                      &        $-$1.49$\,\pm\,$0.14         &        \dots                           \\     
                                                    &             B\dotfill              &                    $-$13.23                                       &                     41.40                               &           \phantom{$-$}0.13                     &      17.73$\,\pm\,$0.02          &             28.50                            &        \phantom{$-$}0.53                      &        $-$1.16$\,\pm\,$0.14         &        \dots                           \\     
                                                    &             C\dotfill              &                    $-$13.76                                       &                     40.82                               &                    $-$0.45                      &      18.30$\,\pm\,$0.03          &             28.22                            &        \phantom{$-$}0.26                      &        $-$1.59$\,\pm\,$0.15         &        \dots                           \\     
                                                    &             D\dotfill              &                    $-$13.73                                       &                     40.67                               &                    $-$0.60                      &      18.01$\,\pm\,$0.02          &             28.10                            &        \phantom{$-$}0.13                      &        $-$1.10$\,\pm\,$0.14         &        \dots                           \\     
\hline                                                                                                                                                                                                                                                                                                                                                                                                                                                                                    
\multicolumn{1}{c}{\multirow{2}{*}{HCG91}}          &             A\dotfill              &                    $-$11.93                                       &                     42.85                               &           \phantom{$-$}1.58                     &      15.25$\,\pm\,$0.02          &             29.25                            &        \phantom{$-$}1.28                      &        $-$0.92$\,\pm\,$0.14         &         AGN                            \\     
                                                    &             C\dotfill              &                    $-$12.33                                       &                     42.18                               &           \phantom{$-$}0.91                     &      16.58$\,\pm\,$0.03          &             28.62                            &        \phantom{$-$}0.65                      &        $-$0.61$\,\pm\,$0.14         &        \dots                           \\     
\hline                                                                                                                                                                                                                                                                                                                                                                                                                                                                                    
\multicolumn{1}{c}{\multirow{3}{*}{HCG96}}          &             A\dotfill              &                    $-$12.30                                       &                     42.63                               &           \phantom{$-$}1.36                     &      15.90$\,\pm\,$0.01          &             29.14                            &        \phantom{$-$}1.17                      &        $-$0.98$\,\pm\,$0.19         &         AGN                            \\     
                                                    &             C\dotfill              &                    $-$13.37                                       &                     41.40                               &           \phantom{$-$}0.13                     &      19.24$\,\pm\,$0.04          &             28.09                            &        \phantom{$-$}0.13                      &        $-$1.65$\,\pm\,$0.16         &        \dots                           \\     
                                                    &             D\dotfill              &                    $-$13.91                                       &                     40.55                               &                    $-$0.72                      &      18.35$\,\pm\,$0.03          &             27.97                            &        \phantom{$-$}0.01                      &        $-$1.49$\,\pm\,$0.14         &         SFN                            \\      
\hline
\end{tabular}
\label{starformationtable}
\end{center}
{\bf Notes:} $^{a}\,$We estimate an upper limit of  0.02$\,$dex for the error in the $\log {F_{{\text{H}}\alpha  + {\text{NII}}}}$ measurements.
             $^{b}\,$We estimate an error of  0.14$\,$dex in the $\log {L_{{\rm{H}}\alpha }}$ and $\log$ SFR$_{{\rm H}\alpha}$ measurements based on the uncertainties of the applied corrections.
             $^{c}\,$We consider an error of  0.20$\,$dex in the $\log {L_{{\rm{FUV}}}}$ and $\log$ SFR$_{{\rm FUV}}$ measurements based on the 0.5mag uncertainty of the dust correction.
             $^{d}\,$Activity according to \citet{martinez10}. SFN -- star-forming nucleus, TO -- transition object, AGN -- active galactic nucleus.
             $^{e}\,$We considered the $g'$$-$$r'$ color criterion of \citet{bell03} to identify blue, star-forming galaxies in this system.
\end{table*}

\subsection{Galaxy stellar masses}
We  have estimated  galaxy stellar  masses by  computing stellar  mass-to-light ratios  in  various passbands  based on  the measured  $g'$$-$$r'$ galaxy  colors. According  to \citet{bell03}  stellar
mass-to-light ratios can be parametrized by $g'$$-$$r'$ colors as

\begin{equation}
\label{stellarmass}
\log {({M_\star}/L)_ \odot } = {a_\lambda } + {b_\lambda } \times (g'-r')
\end{equation}

where the coefficients  $a_\lambda$ and $b_\lambda$ define the relation  for different passbands (see Table 7  in \citet{bell03} for the corresponding coefficients).  We computed stellar mass-to-light
ratios in $g'$, $r'$ and  also in the near-infrared $JHK$ filters since 2MASS $JHK$ magnitudes  are available for most of our galaxies\footnote{We took JHK magnitudes  from SIMBAD when available.}. We
derived galaxy luminosities in $g'$,  $r'$, $J$, $H$, $K$ considering absolute solar magnitudes measured by  C.\ Willmer\footnote{\url{http://mips.as.arizona.edu/~cnaw/sun.html}}. Based on the derived
stellar mass-to-light  ratios and  galaxy luminosities, galaxy  stellar masses were  computed for  every passband. Finally,  we averaged  the derived values  and estimated error  bars as  the standard
deviation of the individual measurements. Stellar masses are shown in Table \ref{photometrictable}.

\subsection{Main sequence of star-forming galaxies}
\begin{figure}
\begin{center}
\includegraphics[width=220pt]{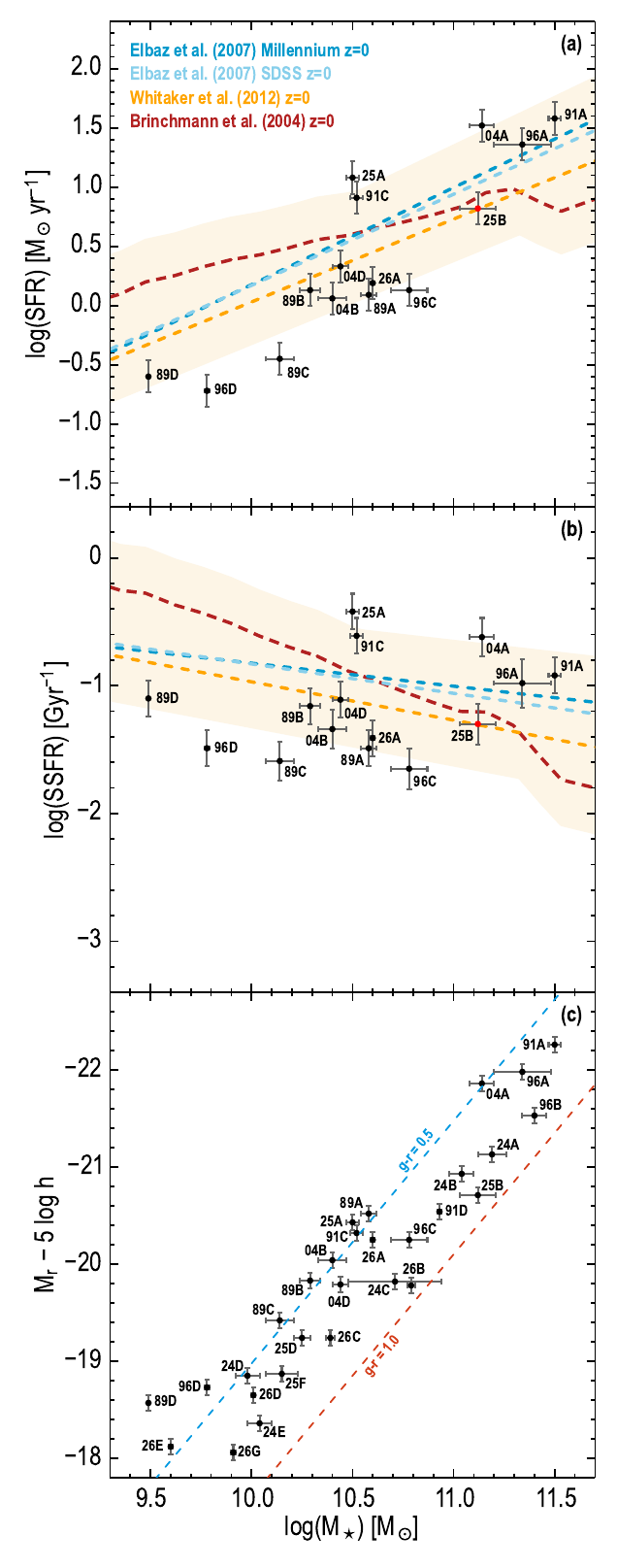}
\end{center}
\caption{\label{mainsequence}\emph{\textbf a)}: $\log$ SFR$-$$\log$ M$_\star$ plane  for all observed HCG member galaxies on the NUV$-r'$ blue sequence. The  dashed lines indicate the main sequence of
star-forming galaxies as shown  in \citet{brinchmann04}, \citet{elbaz07}, and \citet{whitaker12}. The shaded area  gives the $1\sigma$ scatter of these relations. \emph{\textbf  b)}: Same as panel a),
but for  the specific star  formation rate  SSFR. \emph{\textbf c)}:  Absolute magnitude with  respect to  stellar mass. The  dashed lines show  the relations  for galaxies with  $g'$$-$$r'$$=0.5$ and
$g'$$-$$r'$$=1.0$ based on equation \ref{stellarmass}.} 
\end{figure}

It is  now widely known that  star-forming galaxies show  a tight correlation  between their star-formation rates  and stellar masses,  forming a main sequence  of star-forming galaxies in  the $\log$
SFR$-\log M_\star$ plane

\begin{equation}
\log {\rm SFR}[{{\rm{M}}_ \odot }{\rm{\; y}}{{\rm{r}}^{{\rm{ - 1}}}}] =  \alpha \cdot \log {M_\star}[{{\rm{M}}_ \odot }]+\beta
\end{equation}

more massive  galaxies exhibiting  higher star-formation rates.  The relation is  evolving over  cosmic time, with  star-formation rates  reaching maximum values  at $z\sim2-3$ at  the peak  of cosmic
star-formation density and the slope $\alpha$ showing a possible evolution with redshift. In  Fig.\ \ref{mainsequence} we compare our measured star-formation and specific star-formation rates with the
main sequence of local star-forming galaxies (\citealt{brinchmann04}, \citealt{elbaz07}, and \citealt{whitaker12}). The shaded areas show the $1\sigma$ scatter of these relations.

Only galaxies located in  the NUV$-r'$ blue sequence are plotted, except for  HCG25B, which based on its morphology and NUV$-r'$  color is an early-type galaxy (marked with a  red dot) not expected to
show any significant  star-formation, but falls on  the main sequence of  star-forming galaxies. We see  that almost all star-forming  galaxies are located on  the main sequence of  local star-forming
galaxies. Only galaxies HCG89C, HCG96C, and HCG96D are located below the $1\sigma$  thresholds of the various main sequences. Figure \ref{mainsequence} also shows the relation between absolute
$r'$ band magnitudes  and stellar mass. Considering  equation \ref{stellarmass}, the scatter in  the relation results from  the fact that different galaxies  show different $g-r$ colours  and that the
stellar masses shown are the computed averages of stellar mass estimates from various filter passbands.

\subsubsection{Comparison with values from literature}
\begin{figure*}
\begin{center}
\includegraphics[width=\textwidth]{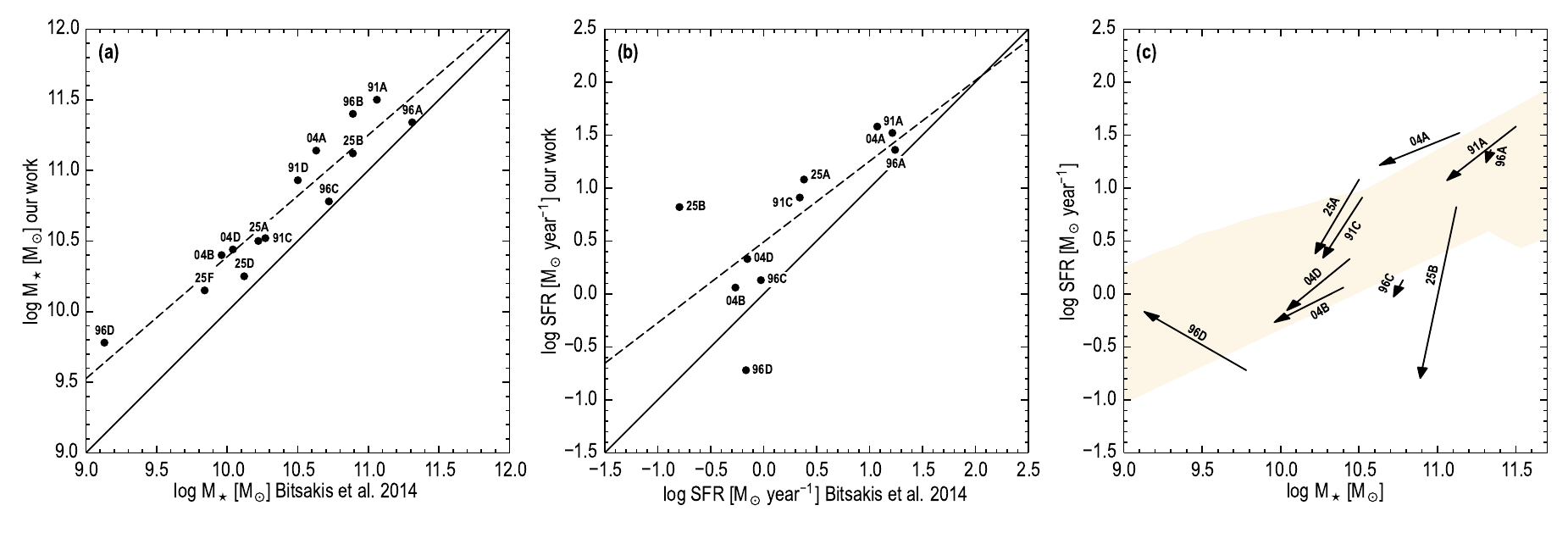}
\end{center}
\caption{\label{bitsakiscomparison} Comparison of  the stellar masses and  H$\alpha$ star-formation rates from  the present work with  \citet{bitsakis14}. \emph{\textbf a)}: Stellar  masses. The solid
line indicates unity while the dashed line is  a linear least squares fit to the data. Stellar masses from our work are systematically above  the \citet{bitsakis14} ones with an average offset of 0.33
in $\log$ M$_\star$. \emph{\textbf b)}: Same as panel \emph{a)} but for star-formation rates. The  star-formation rates from our work are again systematically above the \citet{bitsakis14} ones with an
average offset of 0.46 in $\log\,$SFR. \emph{\textbf  c)}: Arrows show the change of the position of the galaxies on the main  sequence of star-forming galaxies when considering the \citet{bitsakis14}
values instead of ours.}
\end{figure*}

To get  an idea  on the robustness  of our  stellar mass  and star-formation estimates,  we compare  our measurements with  the data  from \citet{bitsakis14}  who study several  of our  galaxies using
multi-wavelength  observations and  SED fitting.  Figure \ref{bitsakiscomparison}  shows the  comparison of  both datasets.  Both stellar  masses and  star-formation rates  from the  present work  are
systematically above the \citet{bitsakis14} values with average offsets of 0.33 in $\log$ M$_\star$  and 0.46 in $\log\,$SFR. However, since \emph{both} parameters are systematically above our values,
datapoints shift  along the  main sequence of  star-forming galaxies  in the $\log$  SFR$-$$\log$ M$_\star$  plane when considering  the \citet{bitsakis14}  values instead of  ours, not  affecting the
interpretation of the present work (see Fig.\ \ref{bitsakiscomparison}c). The only two galaxies where we see a severe mismatch in star-formation rates are HCG25B and HCG96D.

\subsection{Spatial analysis of ionized gas and stellar populations}
To study  the spatial distribution  of the H$\alpha$ emission  tracing the ionized  gas in the  HCG member galaxies, we  created H$\alpha+$[N{\scriptsize II}]  radial surface brightness profiles.  In order to compare  these profiles with  the stellar
component of the galaxies, we also  derived surface brightness profiles in the broadband $g'$ and $r'$ filters. All  surface brightness profiles are shown in Appendix A. We utilized the {\sevensize  ellipse} package for that purpose. The
H$\alpha$ emission in star-forming galaxies is in  general very patchy and irregular making it difficult for {\sevensize ellipse} to  locate the galaxy center and fit ellipses to the flux distribution. To overcome  this problem we fit ellipses in the
much smoother $r'-$band images first, defining  the geometric parameters of all ellipses in these frames and subsequently applying  the same ellipses to the aligned $g'-$band and H$\alpha+$[N{\scriptsize II}] frames. This  way we were able to measure
the azimuthally averaged surface brightness  level in all three frames at the  exact same position of each galaxy, allowing  us to directly compare the surface brightness profiles for  both the stellar and the ionized gaseous  component of the galaxy
disks and reveal any  systematic differences in the corresponding light  distributions. To quantify these differences we define  a light concentration parameter based on the  $\mu_{r'} = 24$ mag arcsec$^{-2}$ isophote,  hereafter $r_{24}$, similar to
\citet{koopmann01}:

\begin{equation}
\label{c30}
{C_{30}} = \frac{{{F_{r'}}(0.3{r_{24}})}}{{{F_{r'}}({r_{24}})}},
\end{equation}

where ${{F_{r'}}({r_{24}})}$  is the total $r'$  band flux within  the $r_{24}$ isophote  and ${{F_{r'}}({0.3r_{24}})}$ the $r'$  band flux within  the $0.3 r_{24}$  isophote. In order to  compare the
optical concentration with the H$\alpha$ emission, we define the H$\alpha$ light concentration parameter analogous:

\begin{equation}
\label{chalpha}
{C_{{\rm{H}}\alpha }} = \frac{{{F_{{\rm{H}}\alpha }}(0.3{r_{24}})}}{{{F_{{\rm{H}}\alpha }}({r_{24}})}}
\end{equation}

where  ${F_{{\rm{H}}\alpha }}({r_{24}})$  is the  total H$\alpha$  flux within  the $r_{24}$  isophote and  ${{F_{{\rm{H}}\alpha }}(0.3{r_{24}})}$  the flux  within the  $0.3 r_{24}$  isophote. Hence,
$C_{{\rm{H}}\alpha}=1$ implies that all H$\alpha$ emission  within $r_{24}$ is located within $0.3 r_{24}$ while $C_{{\rm{H}}\alpha}=0$ indicates that all  the H$\alpha$ emission is located outside of
$0.3 r_{24}$.  Normalizing $C_{{\rm{H}}\alpha}$ by  $C_{30}$ then  allows for a  direct comparison between  the light concentration  in the stellar  and ionized  gaseous components. For  galaxies with
$C_{{\rm{H}}\alpha}/C_{30}>1$ the H$\alpha$ emission is more concentrated than the $r'$ band flux  while galaxies showing values of $C_{{\rm{H}}\alpha}/C_{30}<1$ are more concentrated in the $r'$ band
with respect to  H$\alpha$. Table \ref{lightconcentration} lists $C_{{\rm{H}}\alpha}$  and $C_{{\rm{H}}\alpha}/C_{30}$ for all star-forming galaxies  in the NUV-$r'$ blue sequence.  For those galaxies
hosting an  AGN, we  exclude the  central $\leq0.8$  arcsec emission  in the  H$\alpha$ frames (cf.\  section \ref{agn})  when computing  the H$\alpha$  light concentration  parameter. We  compare our
measurements with the sample of \citet{koopmann04} who analyzed H$\alpha$ profiles of 52 Virgo Cluster spirals.

\begin{figure*}
\begin{center}
\includegraphics[width=\textwidth]{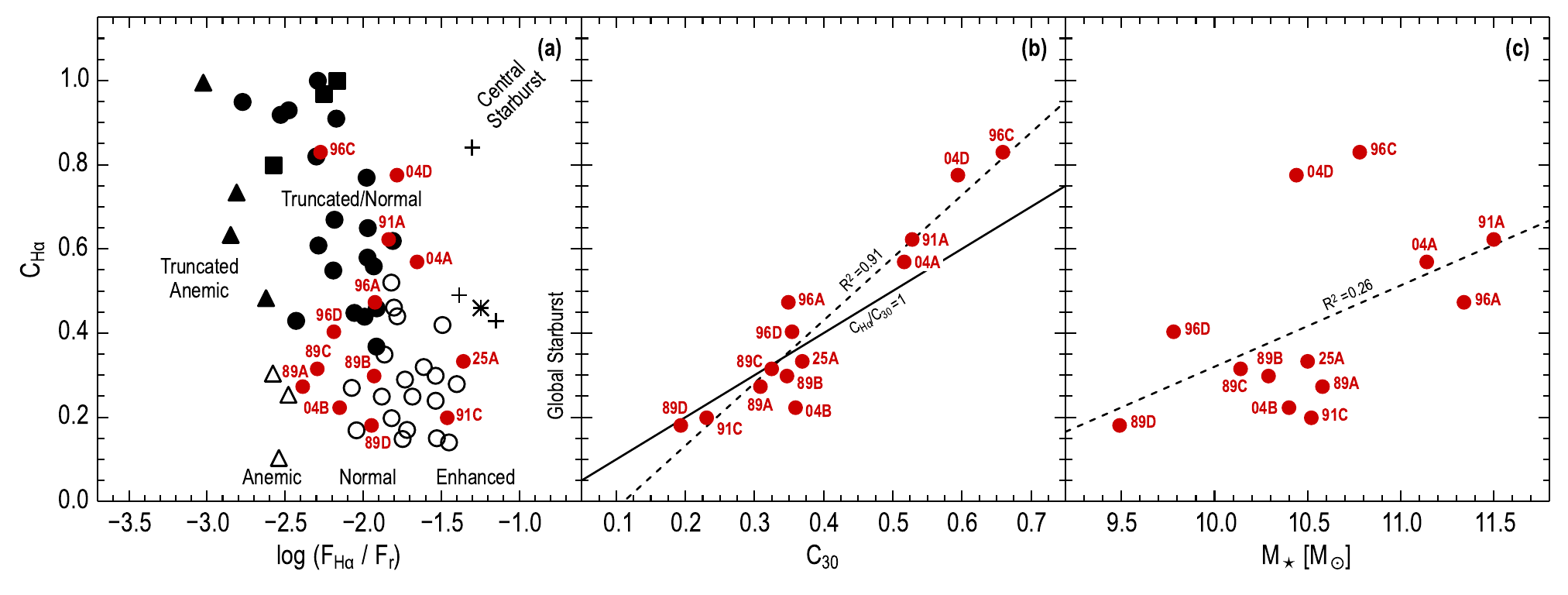}
\end{center}
\caption{\label{truncation}H$\alpha$ light concentration with respect to {\textbf a)}: the NMSFR (i.e.\  log$\,$$f\,$H$\alpha/f_{r'}$), {\textbf b)}: the light concentration in the $r'$ band, {\textbf
c)}: stellar mass.  Black symbols in panel {\textbf a)}  show the 52 Virgo Cluster  spiral galaxies from the study of  \citet{koopmann04}. Full black symbols present galaxies  with truncated H$\alpha$
disks while open ones show normal galaxies. The authors also differentiate between  anemic (triangles), normal (circles), and enhanced (crosses) star-formation (cf.\ Fig.\ 1 \citealt{koopmann04}). Red
circles  show the  star-forming galaxies  from  the present  sample. Panel  {\textbf  b)} shows  that  galaxies that  are compact  in  the $r'$  band, are  even  more concentrated  in H$\alpha$,  i.e.
$C_{{\rm{H}}\alpha}/C_{30}>1$. The dashed line is a linear  least squares fit to the data. The solid line indicates values of $C_{{\rm{H}}\alpha}/C_{30}=1$.  Panel {\textbf c)} shows that more massive
galaxies are more concentrated in H$\alpha$ than less massive ones. The dashed line shows a linear least squares fit to the data.} 
\end{figure*}

\begin{figure*}
\begin{center}
\includegraphics[width=\textwidth]{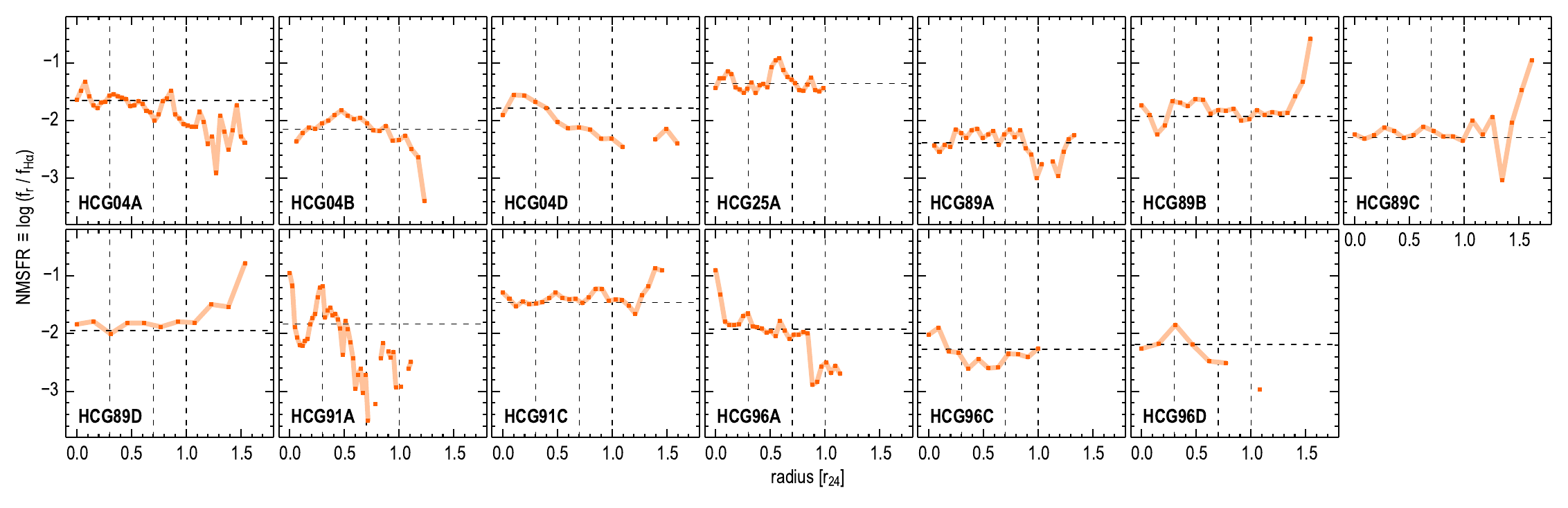}
\end{center}
\caption{\label{truncation1} Radial profiles of normalized  star-formation rate ${\rm{NMSFR}} \equiv \log {\rm{ }}({F_{{\rm{H}}\alpha }}/{F_{r'}})$ for  all galaxies in Table \ref{lightconcentration}.
Radii are given in  units of $r_{24}$. Horizontal dashed lines show  the mean NMSFR estimated from the  integrated ${F_{{\rm{H}}\alpha }}$ and ${F_{r'}}$ fluxes within  the $r_{24}$ isophote. Vertical
dashed lines show radii at $0.3r_{24}$, $0.7r_{24}$, and $1.0r_{24}$.}
\end{figure*}

Figure \ref{truncation}a  shows our derived H$\alpha$  light concentrations with  respect to normalized  star-formation rate (NMSFR),  i.e.\ the ratio  of H$\alpha$ to $r'$  band flux. Galaxies  HCG04A, HCG25A, HCG91C  fall in the region  of enhanced
star-formation in the \citet{koopmann04} diagram,  which is expected since these galaxies are also located  in the uppermost envelope of the main sequence  of star-forming galaxies as shown in Fig.\ \ref{mainsequence}. The  HCG galaxies cover a broad
range  in $C_{{\rm{H}}\alpha}$  from  0.18  to 0.83,  with  galaxies HCG04A,  HCG04D,  HCG91A, HCG96C  exhibiting  values  of $C_{{\rm{H}}\alpha}>0.5$.  Figures  \ref{hcg04}$-$\ref{hcg96} show  surface  brightness profiles  in  $g'$,  $r'$, and  $2.5
\log\,$(H$\alpha+$[N{\scriptsize II}]) for all  galaxies in Table \ref{lightconcentration}. Vertical dashed  lines indicate radii of 0.3 $r_{24}$  and $r_{24}$. We note that for galaxies  HCG04A, HCG91A, and HCG96A the steep  very central increase of
H$\alpha+$[N{\scriptsize II}]  surface brightness is due  to the presence  of an AGN (unresolved,  hence $\leq0.8$ arcsec)  which we exclude  for the computation of  $C_{{\rm{H}}\alpha}$ in these  galaxies. \citet{koopmann04} consider galaxies  to be
truncated in  H$\alpha$ if  the NMSFR  drops significantly  by a  factor of  at least  10, i.e.\  1.0 dex  in the galaxy  outskirts ($0.7r_{24}-1.0r_{24}$).  Figure \ref  {truncation1} shows  radial profiles  of the  NMSFR for  all galaxies  in Table
\ref{lightconcentration}. Horizontal dashed lines  show the mean NMSFR estimated from the integrated  ${F_{{\rm{H}}\alpha }}$ and ${F_{r'}}$ fluxes within the $r_{24}$  isophote as shown in Fig.\ \ref{truncation}. Vertical dashed  lines show radii at
$0.3r_{24}$, $0.7r_{24}$, and $1.0r_{24}$.  We see that HCG04B shows a steep cutoff  in the NMSFR by 0.76 dex outside $r_{24}$  which could be caused by the interaction with  the nearby companion HCG04D. In fact, when looking  at the H$\alpha$ map in
Fig.\ \ref{hcg04},  one can see that  in the southern part  of the galaxy, closer  to the companion HCG04D,  there are hardly any  bright HII regions with  respect to the rest  of the galaxy. We  also note a steep  decline in NMSFR for  HCG96A in the
$0.7r_{24}-1.0r_{24}$ bin by 0.89 dex. The only other galaxies that show a strong negative gradient in NMSFR are HCG89A and  HCG91A. However, in these galaxies the NMSFR rises again towards the outskirts. Based on these measurements and the fact that
\citet{koopmann04} consider a galaxy to be truncated when the NMSFR decreases by a factor of at least 1.0  dex within ($0.7r_{24}-1.0r_{24}$), technically speaking none of our galaxies shows truncation, although some show significant
drops in NMSFR in the galaxy outskirts.

Hence, the high ($C_{{\rm{H}}\alpha}>0.5$)  values derived for some galaxies in our  sample can be rather explained by  above average central star-formation or a central  starburst inducing an overall
steep, homogeneous decline of the H$\alpha$ flux from the  center towards the outskirts rather than a sharp cutoff in the outermost parts. Again, HCG96C is  the only galaxy in our sample that seems to
be special with respect to the other galaxies we observed, exhibiting a comparatively normal  star-formation rate with respect to the Virgo spirals but showing the highest H$\alpha$ concentration from
our sample.  It is interesting to  compare our H$\alpha$  surface brightness profiles  with the HI deficiencies  of the groups  as a whole.  \citet{verdesmontenegro01} proposed a scenario  where least
evolved groups with a low level of interaction have  a lower HI deficiency compared to more evolved groups where multiple tidal tails form and gas has  been removed from the galaxies. Once the gas has
been expelled  it can be  heated or destroyed more  easily, resulting in  HI deficient groups.  \citet{verdesmontenegro01} found a  mean HI deficiency  of $0.40\pm0.07$ in HCGs.  Table 6 lists  the HI
deficiencies from \citet{verdesmontenegro01} for our  sample revealing that all groups show comparatively low HI  deficiencies. This is in agreement with our findings that  we don't see any truncation
in the H$\alpha$ disks, suggesting that the groups are in a relatively  early stage of their evolution. However, we see clear signs of interaction in HCG91 and HCG96, exhibiting
prominent tidal tails. These tails are likely the result of a first encounter in these groups, otherwise we would expect a higher HI gas deficiency in these systems.

Figure \ref{truncation}b  shows that there is  a clear trend  between the $r'$  band light concentration $C_{30}$  and $C_{{\rm{H}}\alpha}$ revealing  that more compact  galaxies in the $r'$  band are
systematically more concentrated in  H$\alpha$, i.e. $C_{{\rm{H}}\alpha}/C_{30}>1$. The difference in the $r'$  band light concentration parameters $C_{30}$ for different galaxies  can be explained by
multiple  effects. First,  the difference  in  the steepness  and overall  shape  of the  surface brightness profiles  in  the $r'$ band  for individual  galaxies  (e.g.\ purely  exponential disks  or
multi-component profiles) can  obviously change the $C_{30}$ parameter  in the sense that galaxy  profiles with a higher S\'{e}rsic index  will exhibit a higher light  concentration parameter. Second,
$C_{30}$ also depends on the bulge-to-disk ratio  $B/T$ of the individual galaxies, i.e.\ early-type spirals showing a comparatively high $B/T$ with  respect to late-type spirals, will also show a
higher $C_{30}$ light concentration parameter. Indeed, from  the visual inspection of the $r'$ band frames of galaxies with $C_{30}>0.5$, we clearly  see that all these galaxies host prominent central
bulges compared to  galaxies with $C_{30}<0.5$. Thus, the observed  trend between $C_{30}$ and $C_{{\rm{H}}\alpha}$  could be explained by the fact  that galaxies with more massive bulges  are able to
more effectively attract  the gas towards the galaxy center,  systematically enhancing star-formation in the center  of the galaxies, hence generating larger  $C_{{\rm{H}}\alpha}/C_{30}$ values. It is
also very interesting to compare galaxies in  Fig.\ \ref{truncation}b with Fig.\ \ref{truncation1}. A purely flat profile in NMSFR will put a  galaxy on the $C_{{\rm{H}}\alpha}/C_{30}=1$ line in Fig.\
\ref{truncation}b. Galaxies with $C_{{\rm{H}}\alpha}/C_{30}>1$ on the other hand show an overall declining trend  of NMSFR with radius, i.e.\ there is systematically more H$\alpha$ flux than $r'$ band
flux in the galaxy center than in the galaxy outskirts, producing the comparatively  high $C_{{\rm{H}}\alpha}/C_{30}$ ratios. Finally, Fig.\ \ref{truncation}c shows a slight trend between stellar mass
and $C_{{\rm{H}}\alpha}$, more  massive galaxies exhibiting a higher  H$\alpha$ light concentration. However, there  are two galaxies, HCG04D and  HCG96C, clearly falling off this  sequence, showing a
much higher H$\alpha$ light concentration  for a given stellar mass. When looking at the corresponding  H$\alpha$ maps in Appendix A one can immediately see that the  H$\alpha$ flux is indeed very
concentrated in these galaxies showing hardly any diffuse distrubtion of HII regions, but one concentrated source.

\begin{table}
\begin{center}
\caption{Light concentration parameters and HI deficiencies.}
\begin{tabular}{ccccc}
\hline                                                                                                                                                                                          
group                                               &          galaxy                          &    $C_{{\rm{H}}\alpha}$  &  $C_{{\rm{H}}\alpha}/C_{30}$$^{a}$ & Def$_{\rm HI}$  \\     
\hline
\multicolumn{1}{c}{\multirow{3}{*}{HCG04}}          &             A$^{b}$\dotfill              &      0.57                &          1.10                      &  \multirow{3}{*}{0.03}  \\     
                                                    &             B\dotfill                    &      0.22                &          0.62                      &   \\     
                                                    &             D\dotfill                    &      0.78                &          1.30                      &   \\     
\hline                                                                                                                                                             
\multicolumn{1}{c}{\multirow{1}{*}{HCG25}}          &             A\dotfill                    &      0.33                &          0.90                      &  \multirow{1}{*}{0.26} \\     
\hline                                                                                                                                                             
\multicolumn{1}{c}{\multirow{4}{*}{HCG89}}          &             A\dotfill                    &      0.27                &          0.88                      &  \multirow{4}{*}{0.04} \\     
                                                    &             B\dotfill                    &      0.30                &          0.86                      &   \\     
                                                    &             C\dotfill                    &      0.32                &          0.97                      &   \\     
                                                    &             D\dotfill                    &      0.18                &          0.93                      &   \\     
\hline                                                                                                                                                             
\multicolumn{1}{c}{\multirow{2}{*}{HCG91}}          &             A$^{b}$\dotfill              &      0.62                &          1.18                      &  \multirow{2}{*}{0.24} \\     
                                                    &             C\dotfill                    &      0.20                &          0.86                      &   \\     
\hline                                                                                                                                                             
\multicolumn{1}{c}{\multirow{3}{*}{HCG96}}          &             A$^{b}$ \dotfill             &      0.47                &          1.36                      &  \multirow{3}{*}{<0.17} \\     
                                                    &             C\dotfill                    &      0.83                &          1.26                      &   \\     
                                                    &             D\dotfill                    &      0.40                &          1.14                      &   \\       
\hline                                                                                                                           
\end{tabular}                                                                                                                  
\label{lightconcentration}                                                                        
\end{center}
{\bf Notes:} $^{a}\,$$C_{{\rm{H}}\alpha}/C_{30}>1$: H$\alpha$  emission is more concentrated  than the corresponding $r'$ band flux. $^{b}\,$Corrected for central AGN.
\end{table}

To further investigate the morphology of the stellar component we also derived $g'$$-$$r'$ color  maps, clearly highlighting the distribution of old and young stellar populations within the HCG member
galaxies. The color maps were created by considering only sources above $1\sigma$ sky noise in each filter.  This way we avoid unreliable results caused by low S/N in any of the two passbands. Figures
\ref{hcg04}$-$\ref{hcg96} show  H$\alpha+$[N{\scriptsize II}]  maps, $g'$$-$$r'$ color  maps, and  the derived  surface brightness profiles for  all HCG  member galaxies located  in the  NUV$-r'$ blue
sequence.

\subsection{Star-forming regions associated with tidal tails\label{tdgsection}}
As noted earlier, HCGs  provide the ideal environment for the birth  of star-forming star clusters and TDGs. Given the  continuous interactions present in these groups, gas  can be removed efficiently
from member galaxies  during group evolution. Once expelled, the  gas can cool, self-gravitate and  form new stars. Examining the  generated H$\alpha+$[N{\scriptsize II}] maps from our  sample, we see
that some groups indeed  show H$\alpha$ knots associated with the tidal  tails seen in these systems. To identify  any star-forming regions we ran {\sevensize SEXTRACTOR}  from \citet{bertin96} on our
H$\alpha+$[N{\scriptsize II}]  maps and considered all  objects above 2$\sigma$ sky  noise and a minimum  extent of 5 pixels  as H$\alpha+$[N{\scriptsize II}] sources.  Table \ref{tdgcandidates} lists
coordinates and H$\alpha+$[N{\scriptsize  II}] fluxes while Figures \ref{gallery2}  and \ref{gallery3} show the corresponding locations  of all detected star-forming regions  associated with prominent
tidal tails in our sample.

In a previous work,  \citet{hunsberger96} reported the detection of TDG candidates in  the tidal tails of HCG26 and HCG96  based on deep $R$ band imaging. However, no  H$\alpha$ observations have been
carried out for these TDG candidates up to  now. The authors mention three TDG candidates in the prominent tidal tail of HCG26 extending towards  the northwest. We unambiguously identify the three TDG
candidates in our deep $r'$ band images but only see one counterpart, HCG26a, in  the corresponding H$\alpha+$[N{\scriptsize II}] map above the detection threshold. Hence, the other two TDG candidates
detected by  \citet{hunsberger96} either do not  show any significant ongoing  star-formation, which  would be expected if  they formed recently  due to the interactions  of the host galaxies,  or the
galaxies are  not group  members but  rather background galaxies  in chance  projection. This  is plausible since  \citet{hunsberger96} did  not confirm the  group membership  of these  TDG candidates
spectroscopically. Although HCG26a shows some emission in  H$\alpha+$[N{\scriptsize II}], its morphological appearance suggests that this galaxy is likely a background  galaxy as well, a spiral with a
prominent bulge seen  edge on. Hence the detected emission  in H$\alpha+$[N{\scriptsize II}] might be  just an artefact due to image  processing. The generated $g'$$-$$r'$ color maps  support the idea
that the 3  galaxies are merely background galaxies  in chance projection since they  are very red compared  to the tidal tail they  are embedded in (see Fig.\ref{gr_gallery}).  We measure $g'$$-$$r'$
colors of  1.07 for HCG26a  and 0.85 for HCG26b.  HCG26c was too  faint and diffuse  to estimate a reliable  colour. Redder colors  are expected for  these objects if they  are located at  much higher
redshift. Eventually, the true nature of these objects can only be revealed via follow-up spectroscopy.

In  HCG96,  \citet{hunsberger96} claim  to  detect  three TDG  candidates  as  well, one  in  the  eastern, two  in  the  northwestern tidal  tail  of  HCG96A. We  only  detect  two subclumps  in  our
H$\alpha+$[N{\scriptsize II}] map  of HCG96, one in  the eastern and one in  the northwestern tail. Unfortunately \citet{hunsberger96}  don't give RA and  DEC coordinates of their detections  so it is
tedious to match with the detected sources from our work.

HCG91A hosts the most prominent and  extended tidal tail in our sample and we detect 10  star-forming regions in H$\alpha+$[N{\scriptsize II}] associated with the diffuse light  of the tail as seen in
the optical. One object is of special interest, HCG91i, located at the very tip of the tail.  HCG91i does not show the highest H$\alpha$ luminosity of the detected star-forming regions in HCG91 but it
is the  most prominent source  associated with  the tidal tail  in the $r'$  band image  given its extent,  the round, compact  morphology, and its  location. We  measure magnitudes of  $r'=20.14$ and
$g'=21.92$ resulting in a $g'$$-$$r'$  color of 1.78. This is comparatively red and  suggests that HCG91i hosts an old stellar population (see  Fig.\ref{gr_gallery}). However, considering the emission
in H$\alpha$, the galaxy also seems to form  stars at the present epoch. Given its red colour, we estimate a comparatively high stellar mass  of log$\,M_\star\sim9.5$ for HCG91i using the equations of
\citet{bell03}. \\

To  estimate the  ionized hydrogen  mass of  all detected  star-forming regions  we first  computed  the H$\alpha$  luminosity of  these sources,  applying all  the corrections  highlighted in  Sect.\
\ref{starformationrates}. Then, we derived the corresponding ionizing photon luminosity $Q(H^0)$ considering the relation

\begin{equation}
\label{ionizingflux}
{Q(H^0)}[{\rm{photons }} {{\rm{\; s}}^{ - 1}}] = 7.31 \times {10^{11}}{L_{{\rm{H}}\alpha }[\rm{erg\; s}^{-1}]}
\end{equation}

from \citet{osterbrockferland}. The ionized hydrogen mass can then be computed via

\begin{equation}
\label{HIImass}
{M_{{\rm{HII}}}}[{{\rm{M}}_ \odot }] = {Q_0}{m_p}n_e^{ - 1}\alpha _B^{ - 1}
\end{equation}

where $m_p$ is  the proton mass, $n_e$ the  electron density, and $\alpha_B$ the  recombination coefficient (\citealt{osterbrockferland}). We assumed an  electron density of $n_e =  400$ cm$^{-3}$ and
$\alpha_B=2.59\times10^{-13}$cm$^{3}$s$^{-1}$. Table \ref{tdgcandidates} lists H$\alpha$ luminosities and the derived ionized hydrogen mass of all detected star-forming regions.

\begin{table*}
\begin{center}
\caption{TDG candidates.}
\begin{tabular}{ccccccccc}
\hline                                                                                                                                                                                          
\multirow{2}{*}{group}               &    \multirow{2}{*}{source}    &  \multirow{2}{*}{$\alpha_{2000}$}   &  \multirow{2}{*}{$\delta_{2000}$}    &     $\log {F_{{\text{H}}\alpha  + {\text{NII}}}}$         &         $\log {L_{{\rm{H}}\alpha }}$     &    $\log$ SFR$_{{\rm H}\alpha}$$^{b}$     &     $\log {M_{{\rm{HII}}}}$     &    distance$^{a}$     \\      
                                     &                               &                                     &                                      &     [erg s$^{-1}$cm$^{-2}$]                               &         [erg s$^{-1}$]                   &    [M$_{\odot}$$\,$year$^{-1}$]           &     [M$_{\odot}$]               &      [kpc]            \\
\hline
\multicolumn{1}{c}{\multirow{3}{*}{HCG26}}     &                 a\dotfill               &                 03 21 54.70                    &                 $-$13 38 34.5                   &                                  $-$15.41                              &                             38.82                  &          $-$2.45                                      &          3.60                             &        17                 \\      
                                               &                 b\dotfill               &                 03 21 54.58                    &                 $-$13 38 25.3                   &                                 $\cdots$                             &                            $\cdots$                &        $\cdots$                                     &        $\cdots$                           &        22                 \\      
                                               &                 c\dotfill               &                 03 21 54.75                    &                 $-$13 38 18.4                   &                                 $\cdots$                             &                            $\cdots$                &        $\cdots$                                     &        $\cdots$                           &        26                 \\      
\hline                                                                                                                                                                                                                                                                                                                                                                                                                                   
\multicolumn{1}{c}{\multirow{10}{*}{HCG91}}    &                 a\dotfill               &                 22 09 04.13                    &                 $-$27 48 08.7                   &                                  $-$15.04                              &                             38.95                  &          $-$2.32                                      &          3.72                             &        26                 \\      
                                               &                 b\dotfill               &                 22 09 03.64                    &                 $-$27 48 17.4                   &                                  $-$15.45                              &                             38.53                  &          $-$2.74                                      &          3.31                             &        27                 \\      
                                               &                 c\dotfill               &                 22 09 09.17                    &                 $-$27 49 13.0                   &                                  $-$14.89                              &                             39.09                  &          $-$2.18                                      &          3.86                             &        20                 \\      
                                               &                 d\dotfill               &                 22 09 09.60                    &                 $-$27 49 13.0                   &                                  $-$14.66                              &                             39.33                  &          $-$1.94                                      &          4.10                             &        22                 \\      
                                               &                 e\dotfill               &                 22 09 11.04                    &                 $-$27 49 12.8                   &                                  $-$14.86                              &                             39.12                  &          $-$2.15                                      &          3.89                             &        28                 \\      
                                               &                 f\dotfill               &                 22 09 13.09                    &                 $-$27 48 01.1                   &                                  $-$15.25                              &                             38.73                  &          $-$2.54                                      &          3.50                             &        38                 \\      
                                               &                 g\dotfill               &                 22 09 14.02                    &                 $-$27 47 54.1                   &                                  $-$15.03                              &                             38.95                  &          $-$2.32                                      &          3.72                             &        45                 \\      
                                               &                 h\dotfill               &                 22 09 10.22                    &                 $-$27 47 05.9                   &                                  $-$15.41                              &                             38.57                  &          $-$2.70                                      &          3.34                             &        46                 \\      
                                               &                 i\dotfill               &                 22 09 06.98                    &                 $-$27 46 42.4                   &                                  $-$14.98                              &                             39.00                  &          $-$2.27                                      &          3.77                             &        54                 \\      
                                               &                 j\dotfill               &                 22 09 06.30                    &                 $-$27 46 35.5                   &                                  $-$14.85                              &                             39.13                  &          $-$2.14                                      &          3.90                             &        58                 \\        
\hline                                                                                                                                                                                                                                                                                                                                                                                                                                    
\multicolumn{1}{c}{\multirow{2}{*}{HCG96}}     &                 a\dotfill               &                 23 27 56.73                    &       \phantom{$-$}08 47 21.0                   &                                  $-$15.66                              &                             38.50                  &          $-$2.77                                      &          3.27                             &        22                 \\      
                                               &                 b\dotfill               &                 23 28 01.89                    &       \phantom{$-$}08 46 58.7                   &                                  $-$15.92                              &                             38.24                  &          $-$2.93                                      &          3.02                             &        46                 \\       
\hline
\end{tabular}
\label{tdgcandidates}
\end{center}
{\bf Notes:} $^{a}\,$Distances measured from the peak of the surface brightness distribution in the brightest group galaxies HCG26A, HCG91A and HCG96A.
\end{table*}

\section{Description of individual groups}

\emph{HCG04} --- \citet{hickson82} listed 5  bright galaxies in this group as members. However,  HCG04C and HCG04E exhibit discordant redshifts indicating that  these galaxies are background galaxies.
Also the group membership  of the Sb spiral HCG04B is doubtful, since  it shows a relative radial velocity of 1074  km s$^{-1}$ with respect to the brightest group  member. Peculiar velocities of this
magnitude are rather expected in galaxy clusters than groups. The group is dominated by a  face-on Sc spiral galaxy (HCG04A) showing two well-defined spiral arms with attached tails exhibiting several
knots in H$\alpha$. The  spiral arms themselves show multiple star-forming regions arranged  like beads on a string \`{a} la \citet{barneshernquist92}.  All spectroscopically confirmed member galaxies
fall on the NUV$-r'$ blue  sequence and are located on the main sequence of  star-forming galaxies, also HCG04D, classified as elliptical by \citet{claudia94}. We  observed H$\alpha$ emission in every
single substructure  found in \citet{hunsberger98}. For  HCG04A detailed kinematic data  is necessary to determine  which HII bright knots  belong to the spiral  stucture and which are  tidally formed
objects. We  find 2 objects  in the $r'$  band which are clearly  offset from the  spiral pattern but  don't detect any  H$\alpha$ or UV  emission for these objects,  indicating that these  are likely
background galaxies. The perturbed  spiral arms in HCG04A are signs of  a previous galaxy interaction process. In case the  interaction partner was another group member of  HCG04, the perturbation was
either createdby  a high-speed encounter of  HCG04B, or by  a slower and  more continuous interaction with  HCG04D. \citet{martinez10} measured  emission line ratios  in the group member  galaxies and
classified HCG04D  as star-forming, while  HCG04A and  HCG04B were classified  as transition objects  between star-forming galaxies  and AGNs. Detailed  kinematic data  of HCG04A together  with galaxy
interaction simulation  with the known properties  of HCG04A, B, and  D will reveal which  galaxy interaction can reproduce  the observed perturbation features.  Another option could be  that a former
interaction partner was fully disrupted and has already  merged with HCG04A, leaving behind the observed tidal tails.

\emph{HCG24}  --- \citet{hickson92}  lists 5  bright galaxies  in this  system.  A tidal  tail extending  from HCG24B  towards  the east  indicates recent  interaction. Group  members show  early-type
morphologies, mainly  S0 galaxies,  all located on  the NUV$-r'$ red  sequence. We  couldn't measure any  FUV emission within  the group  above the detection  limit and any  detection in  H$\alpha$ is
negligible (sSFR$<-2$ Gyr$^{-1}$).

\emph{HCG25} ---  Originally, \citet{hickson82} listed  7 galaxies in this  group, however after  measuring radial velocities  \citet{hickson92} noted that HCG25C,  HCG25E, and HCG25G  show discordant
redshifts, leaving only 4 galaxies as members. Except for  the SBc spiral HCG25A, all galaxies in the group fall on the NUV$-r'$ red sequence. The  most prominent feature in this system is the ongoing
interaction between  HCG25B and HCG25F, clearly  seen as a bridge  of stellar material  connecting both galaxies. The  bridge is made  up of two separate  filaments extending from HCG25B  and blending
together when  approaching HCG25F. We  measure strong H$\alpha$  fluxes for HCG25A  and HCG25B. In  fact, HCG25A is located  at the uppermost  envelope of the  main sequence of  star-forming galaxies.
Interestingly, also HCG25B  shows comparatively strong H$\alpha$  emission despite its red  NUV$-r'$ color and is  located on the main  sequence of star-forming galaxies.  \citet{cluver13} detect warm
molecular hydrogen  in this  galaxy and  classify it  as MOHEG  (MOlecular Hydrogen  Emission-line Galaxy).  HCG25B has  also been  observed spectroscopically  with SDSS  and given  its [N{\scriptsize
II}]/H$\alpha$ emission line  ratio it can be considered as  AGN. However, \citet{cluver13} state that  AGN activity is unlikely to be  responsible for the observed H$_{2}$ enhancement  and that shock
excitation through interaction with the IGM is a  plausible mechanism for producing the observed emission. In the $g'$$-$$r'$ color map we see a  comparatively red, disklike structure in the center of
this galaxy. We don't see any H$\alpha$ emission in the bridge filaments above the detection limit  indicating that the bridge doesn't show any SF activity. However, we note that based on the redshift
of the system and the narrowband filter  transmissivity, the H$\alpha$ detection limit is very low for this system (see Fig.\ \ref{transmission}).  Only spatially resolved spectroscopy will reveal the
ionization mechansism present in this galaxy and shed light on the origin of the strong H$\alpha$+[N{\scriptsize II}] flux measured in the present work.

\emph{HCG26} --- This  compact group is dominated  by an edge-on Scd  spiral galaxy (HCG26A) interacting  with an E0 elliptical (HCG26B)  and shows a tidal  tail extending from the  spiral towards the
northwest. \citet{hickson92} lists a total of 7 spectroscopically confirmed group members. \citet{hunsberger96} have studied  this group in detail in the $R$ band and claim to detect three tidal dwarf
galaxies in the  tail. Since we don't see  any H$\alpha$ emission within the  tail, it could be a  stellar stream made up of a  purely old stellar population as  seen around the MW and  M31. Given the
compactness of the system,  it is unclear to which galaxy  the tidal arm belongs. Based on  its position and alignment it could  have formed from HCG26B, assuming that  HCG26B was moving perpendicular
with respect to the disk of HCG26A. Kinematic data is needed to verify the origin of the tidal feature.

\emph{HCG89} --- This group is made up of four late-type spiral galaxies, all located on the  NUV$-r'$ blue sequence. The group members do not show strong signs of gravitational interaction. HCG89D is
the galaxy with  the lowest stellar mass  and bluest color in our  sample. \citet{coziol04} studied the  emission-line properties of both HCG89A  and HCG89B, classifying both  galaxies as star-forming
galaxies, showing no AGN activity. From our H$\alpha$ measurements we derive that HCG89B and HCG89D fall  on the main sequence of star-forming galaxies while HCG89A and HCG89C both fall below the main
sequence.

\emph{HCG91} ---  \citet{hickson92} lists 4 group  members in this  group, of which only  three, HCG91A, HCG91C,  and HCG91D fall  in our FOV. The  system is dominated  by a SBc face-on  spiral galaxy
(HCG91A) interacting with a  much fainter SB0 companion (HCG91D) and shows a  prominent tidal tail extending from HCG91A to the  east. HCG91A and HCG91C are both located on  the NUV$-r'$ blue sequence
and fall  on the main  sequence of star-forming  galaxies, while HCG91D  doesn't show any  emission in H$\alpha$ or  FUV, falling on  the NUV$-r'$ red  sequence. Although \citet{cluver13}  detect warm
molecular hydrogen in galaxies HCG91A  and HCG91C, which, based on their work, can only be  explained by shock excitation, these shocks do not seem to  suppress star-formation in both galaxies. HCG91A
hosts a prominent tidal  tail clearly visible in our deep $r'$ band  image. The tail splits into two subcomponents  pointing northeast. We find several H$\alpha$ knots along  the tidal tail indicating
active, interaction-induced star-formation. We  report a yet unidentified TDG at the tip  of the tidal tail in HCG91. Given the  low HI deficiency of the system, the tidal  tail very likely originates
from a first encounter in this group. Indeed, \citet{bitsakis14} classify the system as dynamically young.

\emph{HCG96} --- HCG96 is  made up of 4 members, dominated by a  Sc galaxy (HCG96A) interacting with an Sa companion  (HCG96C) and shows two long, filamentary tidal tails  extending from HCG96A to the
east and to the northwest. Based  on the $g'$$-$$r'$ color map for this galaxy, we note that the  the northwestern tail exhibits a much bluer color than the eastern one.  This is in agreement with the
work of \citet{verdesmontenegro97} who have studied the system in detail and showed a similar  result. \citet{cluver13} have found molecular hydrogen $H_2$ in galaxies HCG96A and HCG96C. HCG96C is the
only galaxy from our sample  that is located in the NUV$-r'$ green valley  (see Fig.\ \ref{greenvalley}) and shows the highest H$\alpha$ concentration  within our sample. \citet{torresflores13} showed
that this galaxy also deviates from the baryonic  Tully-Fisher relation. \citet{cluver13} do not classify HCG96C as a MOHEG, hence shocks do not  seem to be the mechanism suppressing star-formation in
this galaxy. Given its vicinity to HCG96A, a likely explanation for the suppressed star-formation in  this galaxy could be the ongoing interaction with the massive nearby galaxy HCG96A. However, we do
not see a sign  of gas disk truncation in this  object (see Fig.\ \ref{truncation1}). Similar to  HCG91, the group shows a comparatively low  HI deficiency and the tidal tails  likely originate from a
first encounter in this group. The system is also classified as dynamically young by \citet{bitsakis14}. \\

\section{Conclusion}
We have  observed a sample of  seven compact groups from  the well-known catalog of  \citet{hickson82} in $g'$,  $r'$, and a narrow-band  filter located at the  redshifted H$\alpha$ line to  study the
star-forming properties  of these groups.  Given the high-density  environment in compact  groups and  the prevalence of  galaxy-galaxy interactions within  these aggregates, star-formation  of member
galaxies is expected to  be strongly affected by the group environment.  In contrast to the galaxy cluster environment,  where the star-forming properties of individual members  are mostly affected by
the overall cluster potential, star-formation  rates of galaxies in compact groups are more  prone to ongoing galaxy-galaxy interactions. Hence, a smooth trend  of star-forming properties with respect
to group  mass is  not expected, whereas  this is  observed in galaxy  clusters.  To test these  assumptions, we investigated  the star-forming  properties of our  sample by  measuring both
integrated star-formation rates of the  galaxies as a whole, and in 2-D, by  generating H$\alpha$ maps and constructing azimuthally averaged H$\alpha$ surface  brightness profiles. Since the gas-disks
of star-forming galaxies are easily distorted by  ongoing galaxy-galaxy intereactions, clear signs of interactions such as truncated H$\alpha$ profiles are then  expected to be observed. We sum up our
findigs as follows:\\

1. Group member galaxies are found in  two pronounced and distinct sequences in the NUV-$r'$ vs. absolute $r'$ band magnitude plane,  unambiguously separating blue star-forming galaxies from passivley
evolving, quiescent red galaxies following the  well-defined blue and red sequences for SDSS galaxies in the  local universe. From a total of 23 bright HCG member galaxies  in our sample (for which we
have NUV data), we find 12 galaxies ($\sim52$\%) located in the NUV-$r'$ blue sequence while we find 10 ($\sim44$\%) located in the red sequence.

2. We find only  one galaxy ($\sim4$\%), HCG96C, located in the  NUV-$r'$ green valley. Given its vicinity to  the brightest group galaxy HCG96A, HCG96C is  undergoing strong galaxy-galaxy interaction
causing star-formation  to be quenched and  consequently moving the galaxy  from the blue  star-forming sequence towards the  red, passive sequence. Interestingly,  this galaxy shows also  the highest
H$\alpha$ concentration within our sample.

3. Except for 3 galaxies all galaxies are  located within the 1$\sigma$ confidence intervals of the main sequence of star-forming galaxies in the local  universe. Only HCG89C, HCG96C, and HCG96D fall below the
main sequence. We do not see any systematic enhancement of the global star-formation rate in star-forming HCG member galaxies.

4. Despite being  located on the NUV-$r'$ red sequence,  we measure a comparatively high  H$\alpha$ star-formation rate for the elliptical  galaxy HCG25B, falling on the main  sequence of star-forming
galaxies. The $g'$$-$$r'$ color map reveals a comparatively red, disk-like structure in the central region of this galaxy.

5. We  get a good  agreement between H$\alpha$  and FUV star-formation  rates for the  brightest star-forming galaxies in  our sample, while  fainter galaxies show  higher star-formation rates  in FUV
compared to H$\alpha$.

6. We find  13 star-forming regions associated with  the tidal tails of HCG26,  HCG91, and HCG96. While we  find only one H$\alpha$ source in  the tail of HCG26 and  2 in the tidal tails  of HCG96, we
detect 10 H$\alpha$ sources along the  tidal tails of HCG91. In particular we find a prominent object,  HCG91i, in both H$\alpha$ and the broadband $r'$ image at the tip  of the extended tidal tail in
HCG91.
 
7. Based on  radial profiles of the  normalized star-formation rate (NMSFR)  and the truncation criterion from  \citet{koopmann04}, none of our  galaxies show H$\alpha$ disk  truncation, although some
galaxies (HCG04B, HCG91A,  HCG96A) show a significant drop  in the NMSFR in the  galaxy outskirts. Given our findings  and the comparatively low HI  deficiencies of the observed groups,  we argue that
these systems did not experience a lot of galaxy-galaxy interactions so far. Groups that show prominent tidal tails (HCG91A, HCG96A) are likely having their first encounter.

8. We find a clear trend that  more massive galaxies show a higher concentration in the azimuthally averaged H$\alpha$ surface brightness profiles  and that galaxies showing a high light concentration
in the $r'$ band show a systematically higher light concentration in H$\alpha$. \\

\section*{Acknowledgments}
This work is  based on observations obtained  at the Southern Astrophysical  Research (SOAR) telescope, which is  a joint project of  the Minist\'{e}rio da Ci\^{e}ncia,  Tecnologia, e Inova\c{c}\~{a}o
(MCTI) da Rep\'{u}blica Federativa do Brasil, the  U.S. National Optical Astronomy Observatory (NOAO), the University of North Carolina at Chapel Hill  (UNC), and Michigan State University (MSU). This
work has made  use of {\sevensize IRAF} which is  distributed by the National Optical  Astronomy Observatories, which are operated by  the Association of Universities for Research  in Astronomy, Inc.,
under cooperative agreement with the National Science Foundation. We want to thank the anonymous referee  who helped to improve the paper. We also want to thank Sergio Torres-Flores and Claudia Mendes
de Oliveira  for helpful discussions.  PE acknowledges  support from FONDECYT  through grant 3130485.  SP acknowledges support  from the European  Research Council  under the European  Union's Seventh
Framework Programme (FP7/2007-2013)/ERC Grant agreement 278594 - Gas Around Galaxies.\\

\appendix

\section{\label{appendix}H$\alpha$ maps, color maps, and surface brightness profiles of HCG member galaxies}

\begin{figure*}
\begin{center}
\includegraphics[width=\textwidth]{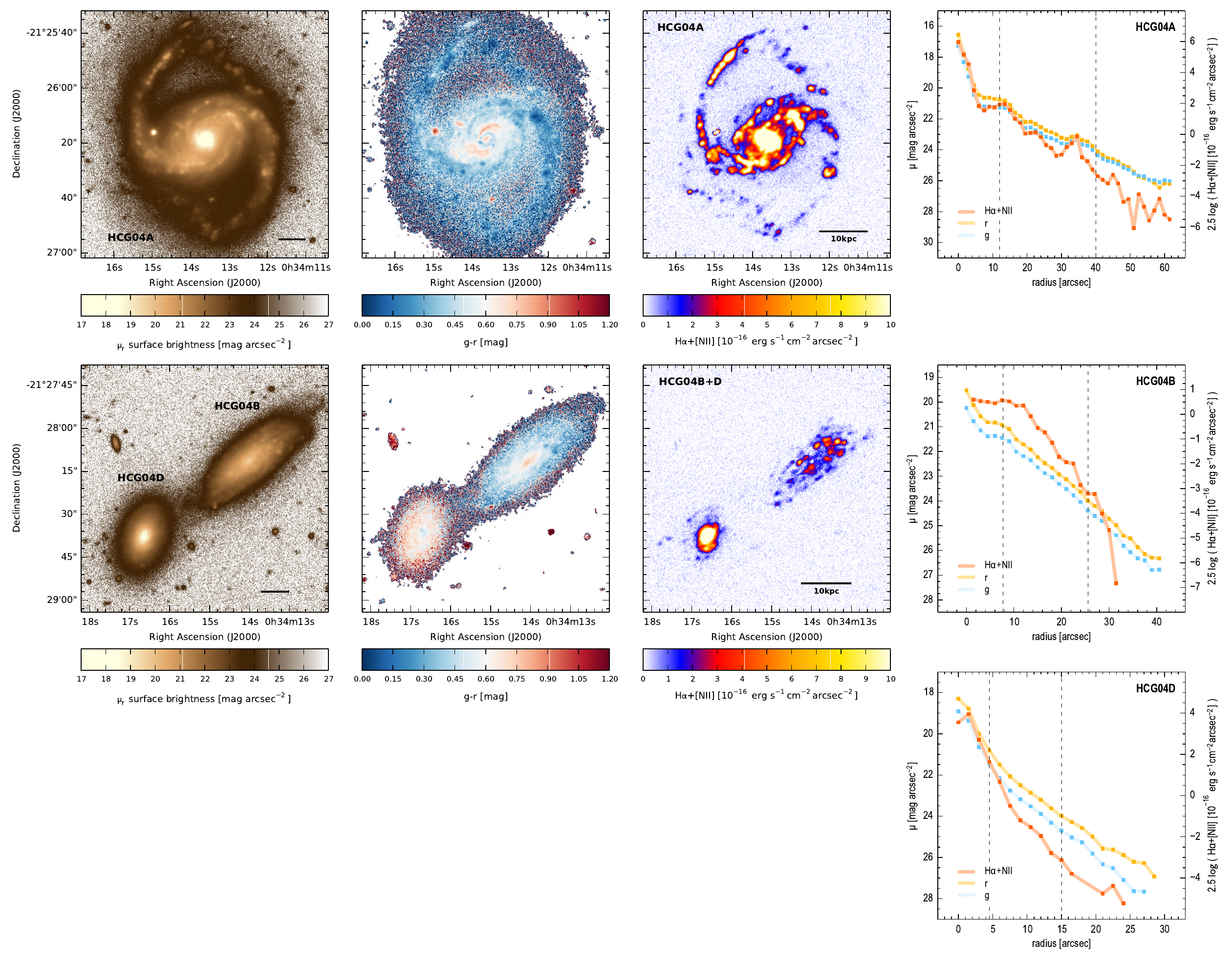}
\end{center}
\caption{\label{hcg04} $r'$ band images, $g'$$-$$r'$ color maps,  H$\alpha+$[N{\scriptsize II}] maps and radial profiles for HCG04A and HCG04B+D. Horizontal scales show  10 arcsec in the $r'$ band and
10kpc in  the H$\alpha+$[N{\scriptsize  II}] maps.  Applying all corrections  and conversion  from Sect.\  \ref{starformationrates}, an H$\alpha+$[N{\scriptsize  II}] flux  of $10^{-16}$  erg s$^{-1}$
cm$^{-2}$ arcsec$^{-2}$ corresponds to a  surface star-formation rate of $\Sigma_{\rm SFR}=6.34\cdot10^{-3} $  M$_\odot$ yr$^{-1}$ kpc$^{-2}$. We note, however, that this  conversion is likely to vary
within the  galaxy. The sky noise  (1$\sigma$) corresponds to $0.20\cdot10^{-16}$  erg s$^{-1}$ cm$^{-2}$  arcsec$^{-2}$. Radial profiles are  shown in surface brightness  for $g'$ and $r'$,  and $2.5
\log\,$(H$\alpha+$[N{\scriptsize II}]) for H$\alpha$. Vertical dashed lines indicate radii of 0.3 $r_{24}$ and $r_{24}$ (see text).}
\end{figure*}

\begin{figure*}
\begin{center}
\includegraphics[width=\textwidth]{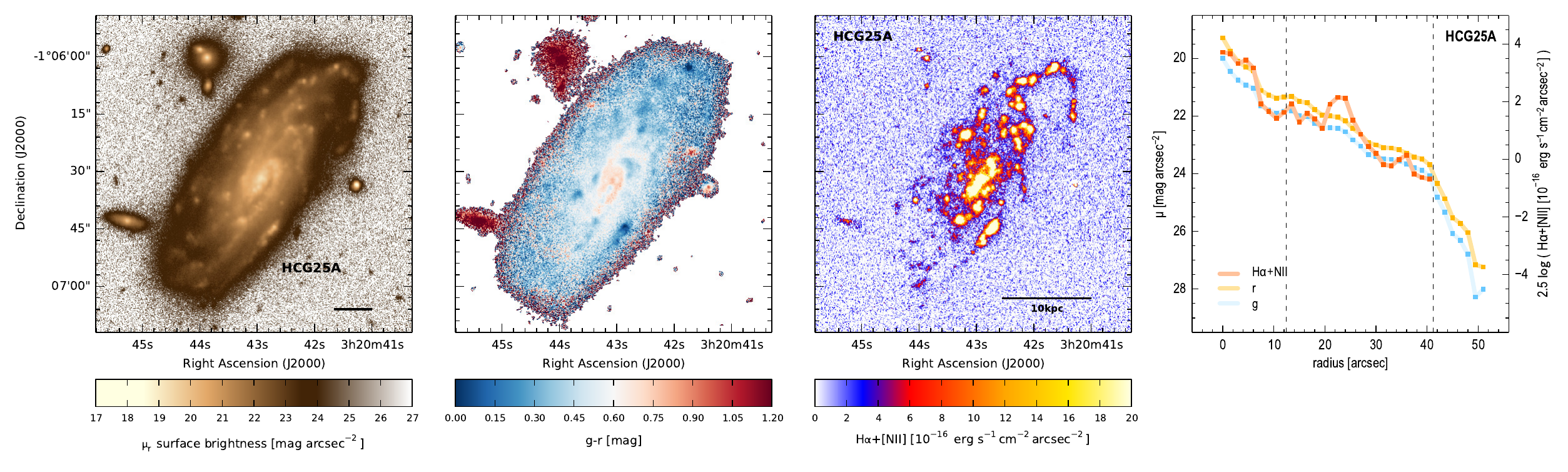}
\end{center}
\caption{\label{hcg25}Same  as  Fig.\ \ref{hcg04}  but  for  HCG25A. A  flux  of  $10^{-16}$  erg s$^{-1}$  cm$^{-2}$  arcsec$^{-2}$  corresponds  to a  surface  star  formation rate  of  $\Sigma_{\rm
SFR}=6.76\cdot10^{-3} $ M$_\odot$ yr$^{-1}$ kpc$^{-2}$. The sky noise (1$\sigma$) corresponds to $1.56\cdot10^{-16}$ erg s$^{-1}$ cm$^{-2}$ arcsec$^{-2}$.}
\end{figure*}

\begin{figure*}
\begin{center}
\includegraphics[width=\textwidth]{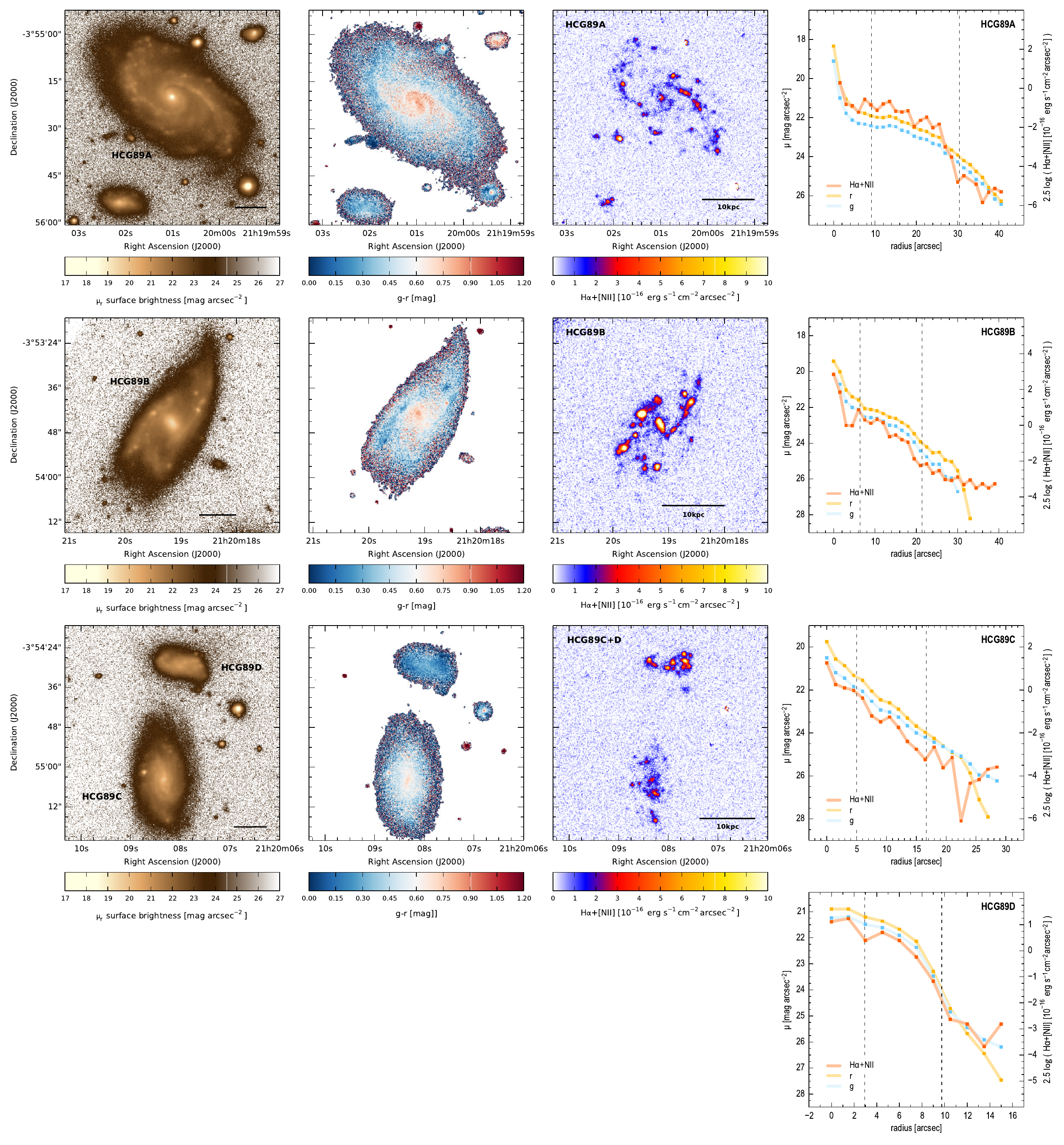}
\end{center}
\caption{\label{hcg89}Same as  Fig.\ \ref{hcg04} but  for HCG89A,  HCG89C+D, and HCG89B.  A flux of  $10^{-16}$ erg s$^{-1}$  cm$^{-2}$ arcsec$^{-2}$ corresponds  to a  surface star-formation  rate of
$\Sigma_{\rm SFR}=6.88\cdot10^{-3} $ M$_\odot$ yr$^{-1}$ kpc$^{-2}$. The sky noise (1$\sigma$) corresponds to $0.37\cdot10^{-16}$ erg s$^{-1}$ cm$^{-2}$ arcsec$^{-2}$.}
\end{figure*}

\begin{figure*}
\begin{center}
\includegraphics[width=\textwidth]{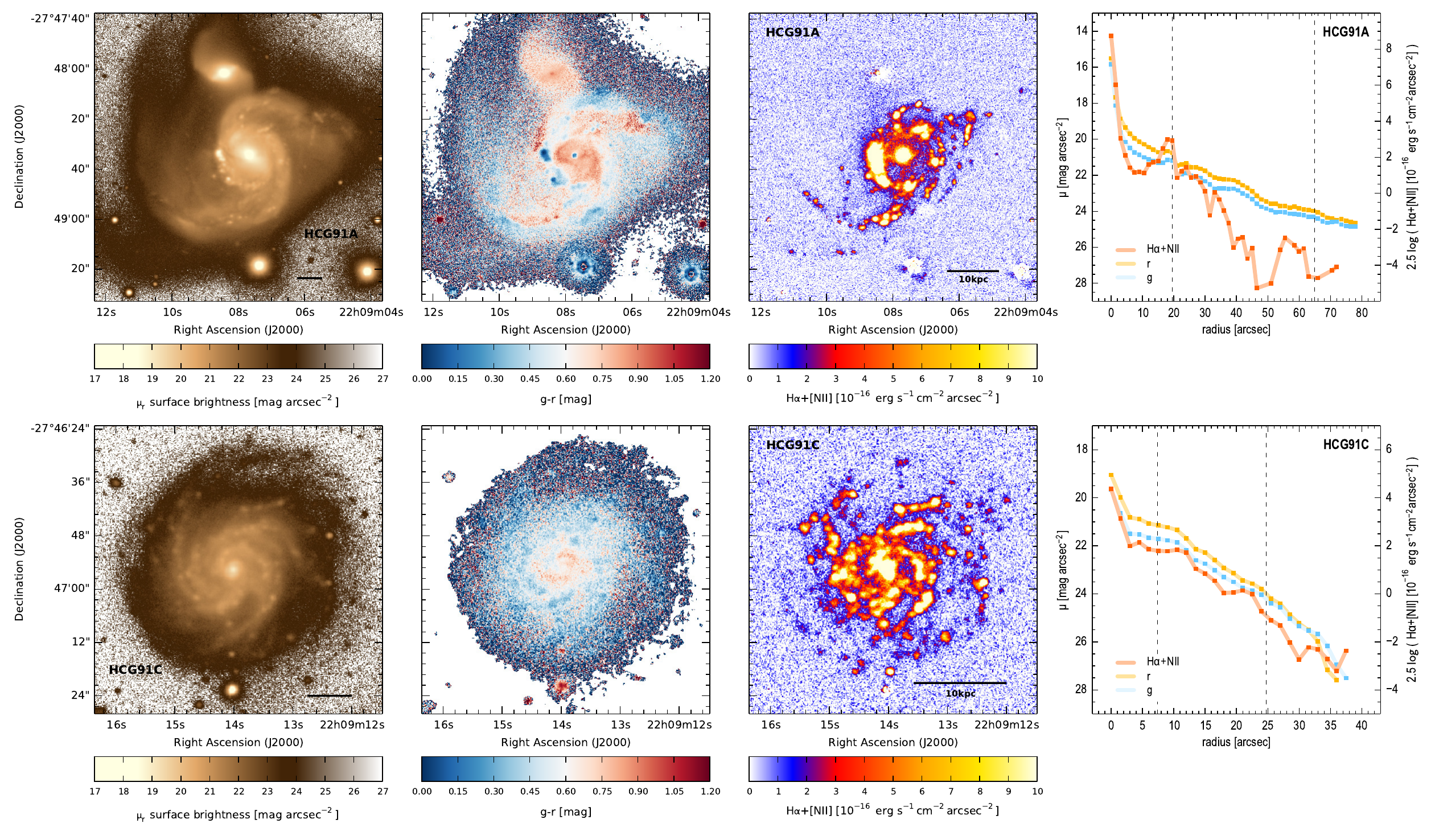}
\end{center}
\caption{\label{hcg91}Same as  Fig.\ \ref{hcg04} but for  HCG91A and HCG91C.  A flux of $10^{-16}$  erg s$^{-1}$ cm$^{-2}$ arcsec$^{-2}$  corresponds to a  surface star-formation rate  of $\Sigma_{\rm
SFR}=6.20\cdot10^{-3} $ M$_\odot$ yr$^{-1}$ kpc$^{-2}$. The sky noise (1$\sigma$) corresponds to $0.62\cdot10^{-16}$ erg s$^{-1}$ cm$^{-2}$ arcsec$^{-2}$.}
\end{figure*}

\begin{figure*}
\begin{center}
\includegraphics[width=\textwidth]{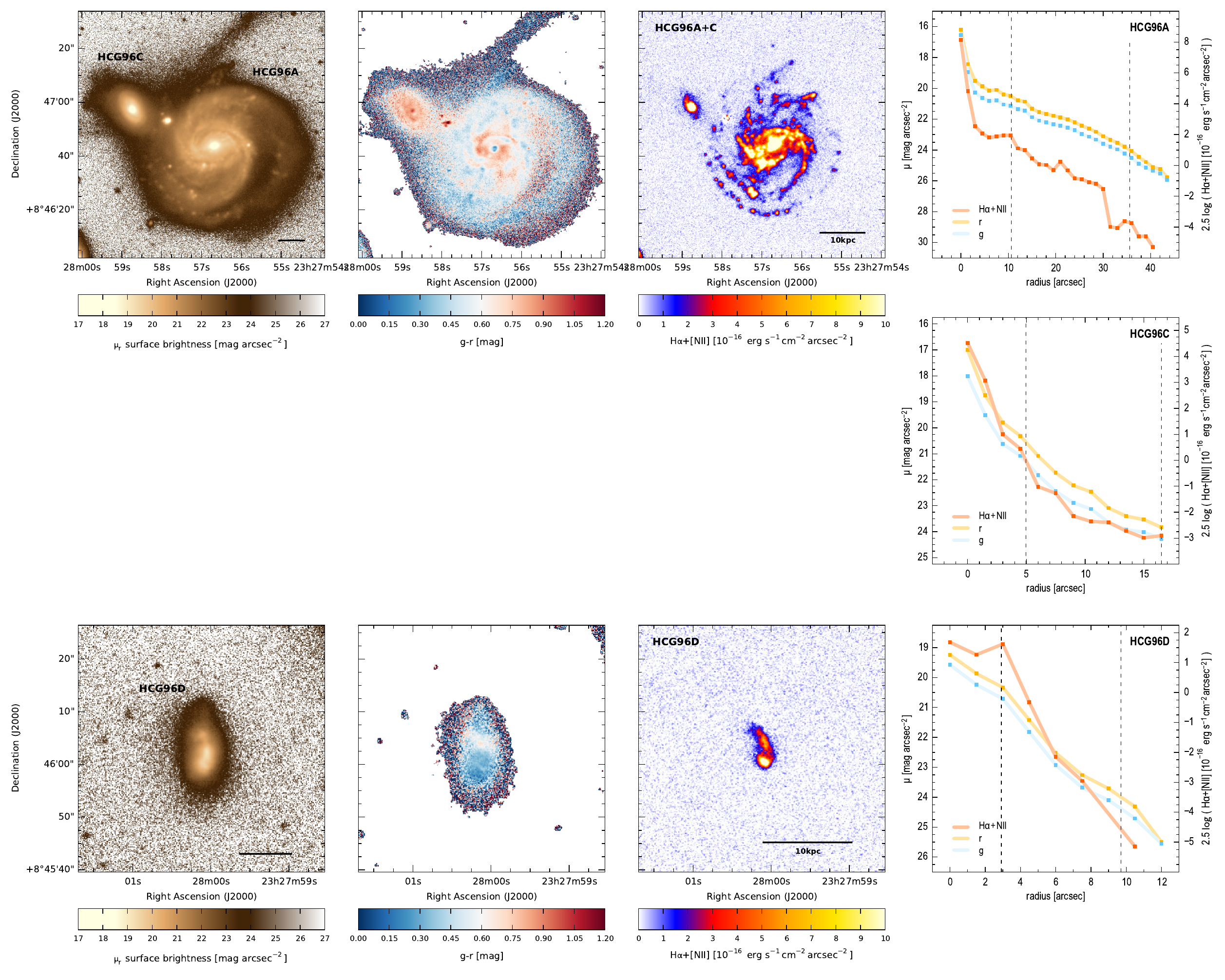}
\end{center}
\caption{\label{hcg96}Same  as Fig.\  \ref{hcg04} but  for  HCG96A+C+D. A  flux of  $10^{-16}$  erg s$^{-1}$  cm$^{-2}$  arcsec$^{-2}$ corresponds  to a  surface  star-formation  rate of  $\Sigma_{\rm
SFR}=6.87\cdot10^{-3} $ M$_\odot$ yr$^{-1}$ kpc$^{-2}$. The sky noise (1$\sigma$) corresponds to $0.28\cdot10^{-16}$ erg s$^{-1}$ cm$^{-2}$ arcsec$^{-2}$.}
\end{figure*}

\section{Color maps for the groups as a whole}

\begin{figure*}
\begin{sideways}
\begin{minipage}{24cm}
\begin{center}
\includegraphics[width=\textwidth]{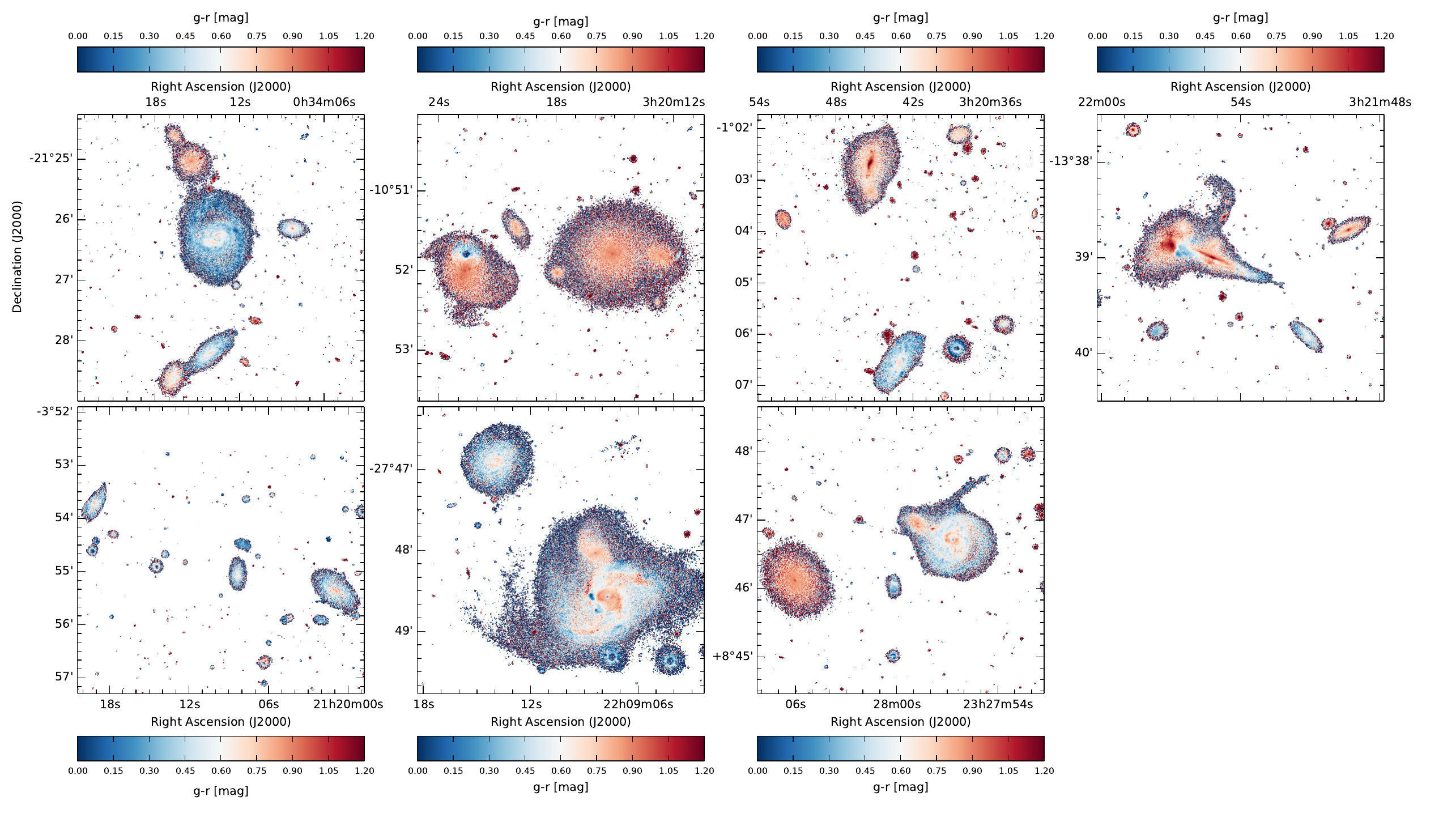}
\caption{\label{gr_gallery} $g'$$-$$r'$ color maps for all observed HCGs. Only sources above 1$\sigma$ sky noise in both filters are shown.}
\end{center}
\end{minipage}
\end{sideways}
\end{figure*}

\label{lastpage}

\end{document}